\documentclass[twocolumn,showpacs,prl,nofootinbib,superscriptaddress,amsmath,amssymb,floatfix]{revtex4-1}

\usepackage{graphicx,color}
\usepackage{dcolumn}
\usepackage{bm}
\usepackage{lineno, soul, ulem}
\usepackage{hyperref}

\usepackage{amsmath}
\usepackage{mathtools}
\usepackage{multirow}

\usepackage{aas_macros}
\usepackage{xcolor}
\usepackage{comment}
\definecolor{darkcyan}{rgb}{0.0, 0.55, 0.55}

\newcommand{\Fermi}{{\it Fermi}}

\newcommand{\SF}{\texttt{skyFACT}}
\newcommand{\Opp}{\texttt{1pPDF}}

\newcommand{\dnds}{$dN/dS$}
\newcommand{\fluxunits}{ph cm$^{-2}$ s$^{-1}$ }

\newcommand{\SFnoGCE}{\texttt{sF-noGCE}}
\newcommand{\SFBBn}{\texttt{sF-B}}
\newcommand{\SFNFWn}{\texttt{sF-NFW126}}

\newcommand{\OppnoGCE}{\texttt{1pPDF-noGCE}}
\newcommand{\OppB}{\texttt{1pPDF-B}}
\newcommand{\OppNFW}{\texttt{1pPDF-NFW126}}

\def \aap  {A\&A}

\def \apj  {ApJ}
\def \apjs  {ApJS}
\def \apjl  {ApJL}

\def \mnras {MNRAS}

\def \prl {PRL}

\begin{document}

\begin{flushleft}
LAPTH-008/21, TTK-21-06
\end{flushleft}

\title{
Dissecting the Inner Galaxy with $\gamma$-Ray Pixel Count Statistics
}

\author{F. Calore}
\email{calore@lapth.cnrs.fr}
\affiliation{Univ. Grenoble Alpes, USMB, CNRS, LAPTh, F-74940 Annecy, France}
\author{F. Donato}
\email{donato@to.infn.it}
\affiliation{Dipartimento di Fisica, Universit\`a di Torino, via P. Giuria, 1, I-10125 Torino, Italy}
\affiliation{Istituto Nazionale di Fisica Nucleare, Sezione di Torino, via P. Giuria, 1, I-10125 Torino, Italy}
\author{S. Manconi}
\email{manconi@physik.rwth-aachen.de}
\affiliation{Institute for Theoretical Particle Physics and Cosmology, RWTH Aachen University, Sommerfeldstr.\ 16, 52056 Aachen, Germany}

\begin{abstract}
 We combine adaptive template fitting and pixel count statistics in order to assess the nature of the Galactic center excess in \Fermi-LAT data. 
 We reconstruct the flux distribution of point sources  well below the \Fermi-LAT detection threshold, and measure their radial and longitudinal profiles in the inner Galaxy. 
 We find that all point sources $and$ the bulge-correlated diffuse emission
each contributes $\mathcal{O}$(10\%) of the total inner Galaxy emission, and disclose a potential sub-threshold point-source contribution to the Galactic center excess.

\end{abstract}

\maketitle

\textit{Introduction.}
The Galactic center excess (GCE) shows up as an unexpected $\gamma$-ray component in the data of the Large Area Telescope (LAT), aboard the \Fermi~satellite, at GeV energies,  
from the inner degrees of the Galaxy \cite{Abazajian:2012pn,Gordon:2013vta,Calore:2014xka,Daylan:2014rsa,TheFermi-LAT:2015kwa}. 
Despite the great interest raised by the GCE discovery, its nature is still  unknown. While the GCE morphology has been found to be consistent with a Navarro, Frenk and White (NFW) profile~\cite{Navarro:1996gj} for annihilating particle dark matter (DM) in~\cite{Abazajian:2012pn,Daylan:2014rsa,Calore:2014nla,Agrawal:2014oha,Murgia:2020dzu,DiMauro:2021raz},
it could also be due to a population of  millisecond pulsars, as proposed by~\cite{Abazajian:2010zy}. Stellar distributions were used as tracers of point sources (PS) emitting below threshold, and turned out to match the morphological features of GCE photons better than DM-inspired templates in~\cite{Bartels_bulgelum, Macias:2016nev, Macias:2019omb}.
All these results were obtained by $\gamma$-ray analyses based on the, so-called, template fitting.
In parallel, complementary methods, based on photon-count statistics and aimed at detecting new point sources below the threshold of the \Fermi~catalogs were developed. They initially 
revealed that the GCE can be entirely due to a population of PS~\cite{Bartels:2015aea,Lee:2015fea}. More recently, the DM interpretation was brought back by~\cite{Leane:2019xiy}, although hampered by systematics affecting photon-count statistical methods~\cite{Leane:2020nmi,Chang:2019ars,Buschmann:2020adf,Zhong:2019ycb,Leane:2020pfc}. 
Techniques involving neural networks have also been explored~\cite{Caron:2017udl,List:2020mzd}. 
As a conclusive probe of the PS nature of the GCE, a fully  multiwavelength approach has been proposed, from radio to gravitational wave observations~\cite{Calore_radioprospects_msp, Calore_GeVexcessGW, Berteaud:2020zef}.
A major limitation to all these studies is the modeling of the Galactic diffuse foreground, and the impact of residual mis-modeled emission on the results' robustness.
As for template fitting methods, the analysis of the  diffuse emission has been recently approached with the \SF~algorithm, which fits the $\gamma$-ray sky by combining methods of image reconstruction and adaptive spatio-spectral template regression~\cite{Storm:2017arh}. The \SF~method has been tested in the {\it Inner Galaxy} (IG) region, and probed to be efficient in the removal of most residual emission for a robust assessment of the GCE properties~\cite{Storm:2017arh,Bartels_bulgelum}. 
Another source of uncertainty  
is the contribution of sub-threshold PSs. Photon-count statistical methods can discriminate  photons from $\gamma$-ray sources  based on their statistical properties~\cite{2011ApJ...738..181M}. 
In particular, the 1-point probability distribution function method \cite{Zechlin:1} (\Opp) fits the contribution of  diffuse and PS components to the $\gamma$-ray 1-point fluctuations histogram. Employing \Opp~on \Fermi-LAT data, it was possible to measure the PS count distribution per unit flux, \dnds, below the LAT detection threshold at high latitudes~\cite{Zechlin:1,Zechlin:2,Manconi:2019ynl}, and to set competitive bounds on DM~\cite{Zechlin:3}.

The scope of this \textit{Letter} is to apply the \Opp~method to \Fermi-LAT data from the IG to understand the role of faint PS to the GCE, while minimizing the mis-modelling of diffuse emission components.
To this end, we adopt a hybrid 
approach which combines, for the first time, adaptive template fitting methods as implemented in \SF, and \Opp~techniques. %

\medskip

\textit{Rationale, data and methodology.} 
We follow  a two-step procedure: 
First, we fit $\gamma$-ray data with
\SF~in order to build a model for the  emission in the region of interest (ROI), maximally reducing residuals found to bias photon-count statistical methods~\cite{Buschmann:2020adf}.  
Secondly, we run  \Opp~fits with  \SF-optimized diffuse models as input,
and assess the role of PS to the GCE.

We analyze 639 weeks of \texttt{P8R3 ULTRACLEANVETO} \Fermi-LAT data~\cite{fermidata} until 2020-08-27. For the \SF~fit, we consider an ROI of 40$^\circ \times 40^\circ$ around the GC~\cite{SFfootnote}, and the $0.3-300$~GeV energy range. 
We closely follow~\cite{Bartels_bulgelum} and update the analysis 
for the increased data set and  4FGL catalog~\cite{Fermi-LAT:2019yla}.
The emission model includes $\gamma$ rays from inverse Compton scattering, $\pi^0$ decay, 4FGL point-like and extended sources, the \Fermi~bubbles, the isotropic $\gamma$-ray background (IGRB), and  the GCE. 
For the latter, we consider a template for the Galactic bulge emission as in~\cite{Bartels_bulgelum}, and one for 
a generalized NFW DM distribution with slope 1.26 (NFW126)~\cite{Calore:2014xka,Daylan:2014rsa}.
We refer to~\cite{supplmat} for more details.

We operate the \Opp~analysis in the energy range $2-5$ GeV \cite{Zechlin:2,Zechlin:3}, restricting to events with best angular reconstruction (evtype=PSF3) and coming from the inner $20^\circ \times 20^\circ$, IG ROI hereafter.
We cut at latitudes $|b| >0.5^\circ$ or $ 2^\circ$ to check the stability of \Opp~results. 
The \Opp-fit model components are: An IGRB template (free normalization), 
a diffuse emission template (free normalization), and an isotropic PS (IPS)\cite{spatialcomment}  population with \dnds~defined by a multiple broken power law: 
\begin{equation}\label{eq:mbpl}
\frac{\mathrm{d}N}{\mathrm{d}S} = A_{\rm S} \cdot 
\begin{cases} 
\left( \frac{S}{S_0} \right)^{-n_1} \;\;\;\; S > S_{\mathrm{b}1}  \, ;\\
\left( \frac{S_{\mathrm{b}1}}{S_0} \right)^{-n_1+n_2} \left( \frac{S}{S_0} \right)^{-n_2}  \;\;\;\; S_{\mathrm{b}2} < S \leq S_{\mathrm{b}1} \, ;
\\
\vdotswithin{\left(\frac{S_{\mathrm{b}1}}{S_0}\right)} & 
\\
\left( \frac{S_{\mathrm{b}1}}{S_0} \right)^{-n_1+n_2} \left( \frac{S_{\mathrm{b}2}}{S_0} \right)^{-n_2+n_3} \cdots \ \left( \frac{S}{S_0} \right)^{-n_{N_\mathrm{b}+1}}  \\  \hspace{3.8cm} S \leq S_{\mathrm{b}N_\mathrm{b}}.\\
\end{cases}
\end{equation}
The free parameters are $A_{\rm S}$, the flux break positions, and the broken power-law indices, $n_i$ \cite{supplmat}.
The IPS \dnds~measured by the \Opp~fit
should recover the \dnds~of \Fermi-LAT detected PS in the bright regime while pushing the PS detection threshold down to lower fluxes~\cite{Zechlin:1,Zechlin:2}. 

Our goal being to quantify the role of PS to the GCE  within the \Opp, we add a \textit{GCE smooth template}  in the \Opp~fit.
As a baseline, we use the best-fit \SF~bulge template  in the \Opp~fit (\OppB), and we define the \SFBBn~diffuse model as the sum of best-fit inverse Compton, $\pi^0$ decay,  \Fermi~bubbles, and extended sources, thus subtracting the bulge emission. The normalization, A$_{\rm B/NFW126}$ for the bulge/NFW126 template, refers to the rescaling factor relative to the best-fit normalization from \SF.

On the one hand, the use of \SF~best-fit diffuse model guarantees a robust characterization of GCE spectrum and morphology against systematics related to the mis-modeling of the diffuse emission~\cite{Buschmann:2020adf,Calore:2014xka}, resolving over/under-subtraction issues by including a large number of nuisance parameters.
The limitations of such a systematic uncertainty are indeed also relevant for the reconstruction of faint PS with \Opp~methods \cite{supplmat}.
On the other hand, the \SF~optimization procedure  mitigates possible systematics related to the mis-modeling of unaccounted components~\cite{Leane:2020nmi}, by allowing spatial re-modulation in the fit templates. 
Also, we stress that this is the first time the stellar distribution in the Galactic bulge as tracer of  GCE photons is used in pixel count statistical analyses (except for brief cross-checks, as in
e.g.~\cite{Leane:2019xiy}).

Besides the bulge, we also consider NFW126 as smooth GCE in
the \Opp~analysis (\OppNFW). In this case, we construct the corresponding 
\SF-optimized diffuse model (\SFNFWn) from 
the \SF~run adopting NFW126 as GCE, in analogy with the \SFBBn~model. 
Such a procedure guarantees maximal consistency between GCE and diffuse models adopted as input in the \Opp.
Finally, to bracket the uncertainties related to the optimization of the diffuse model, we also build a \SF-optimized diffuse template from the \SF~run not including any GCE additional template (\SFnoGCE).

\medskip

{\it Results.}
 Our updated analysis of \Fermi-LAT data with \SF~confirms previous findings from~\cite{Bartels_bulgelum,Macias:2016nev,Macias:2019omb}. 
 A bulge distribution for GCE photons is strongly preferred by data on top of the NFW126-only model ($\sim 10\sigma$), and there is mild  evidence for an additional NFW126 contribution on top of the bulge-only model ($\sim 4\sigma$), cf.~\cite{supplmat}.
This implies that the model maximally reducing the residuals is the \SF~best-fit of the 
run with the bulge.

\begin{figure*}[t]
  \includegraphics[width=0.49\textwidth]{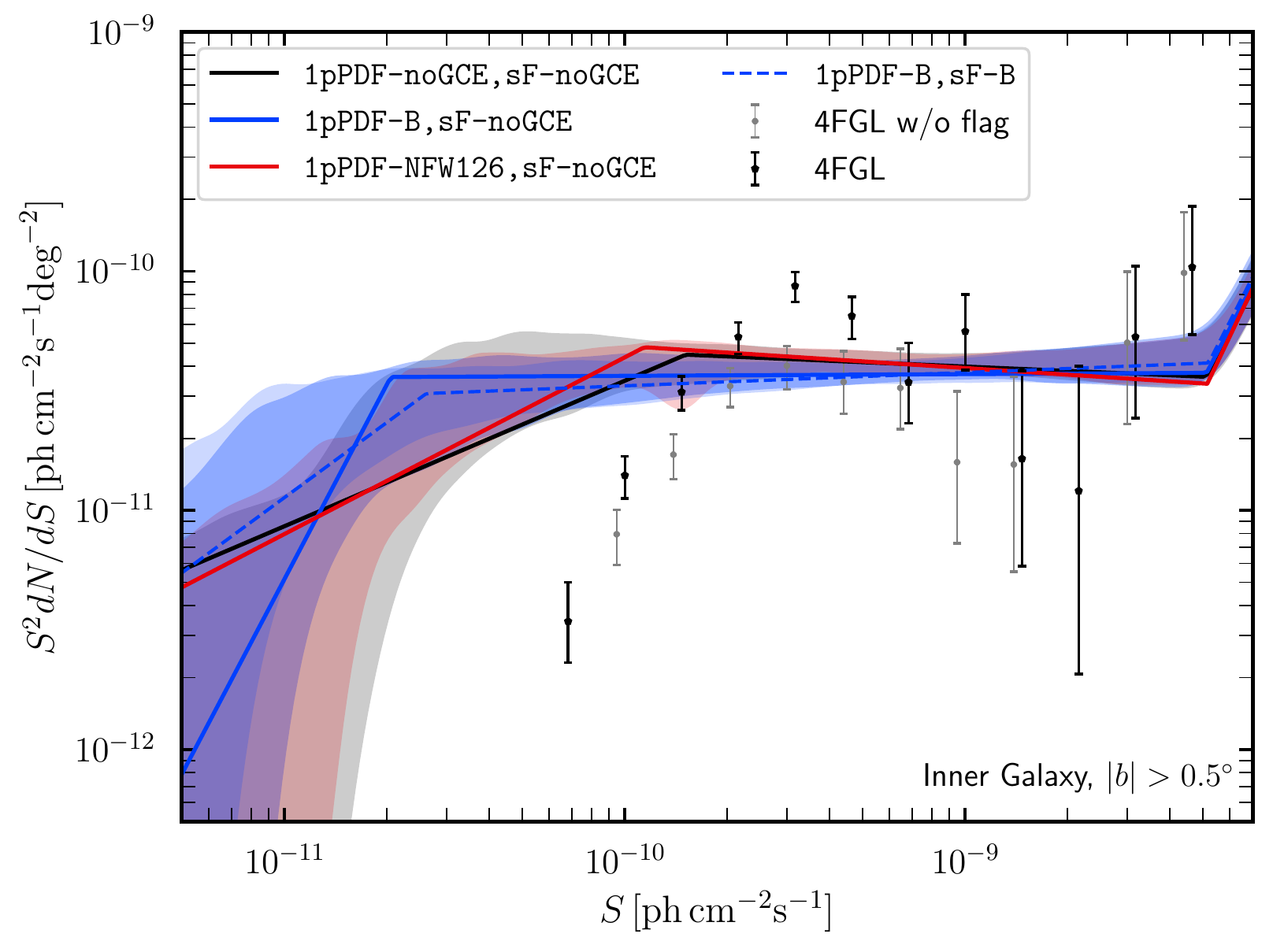}
    \includegraphics[width=0.49\textwidth]{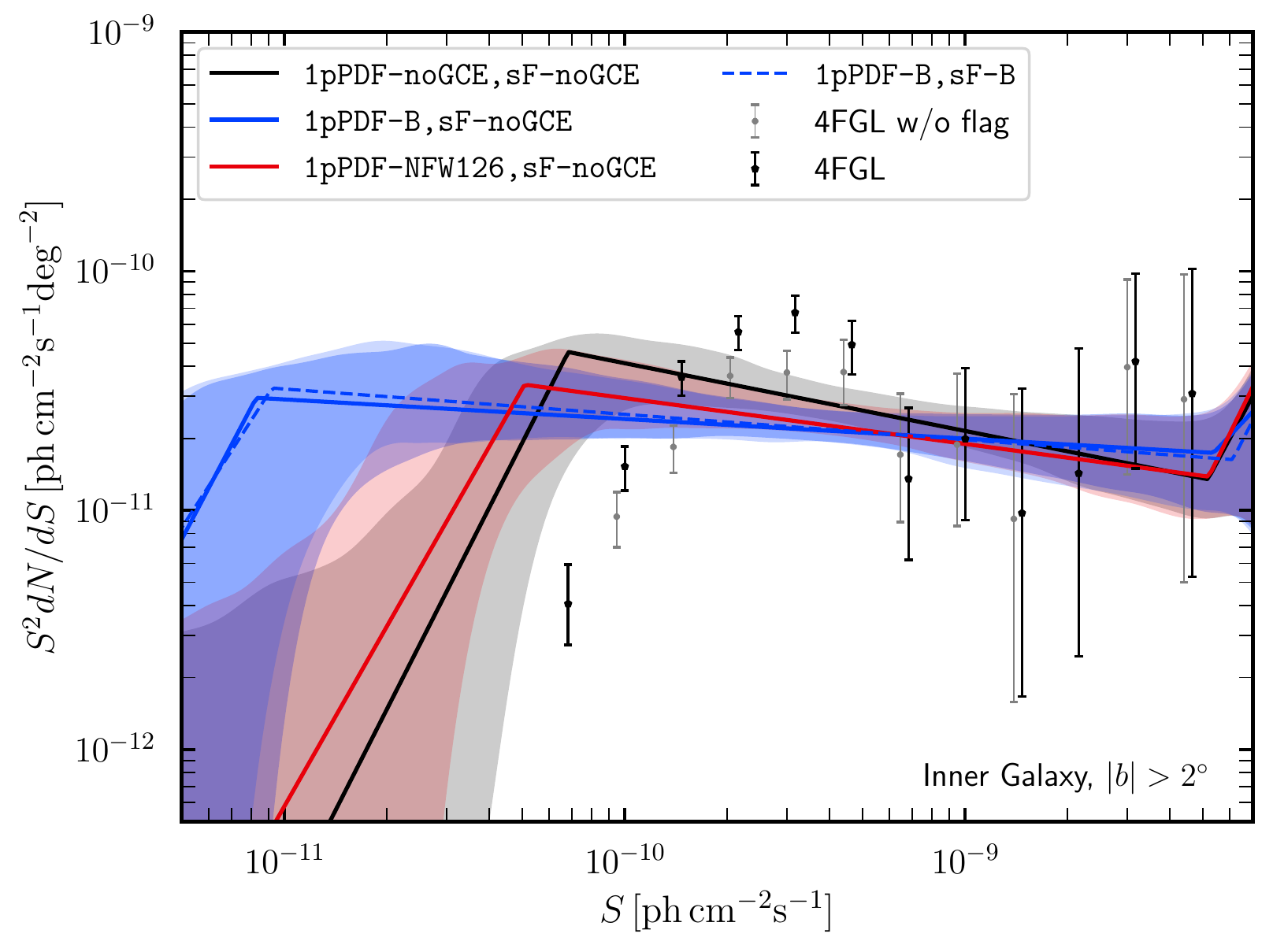}
  \caption{
  \textit{IPS source count distribution in the IG ROI}  from the \Opp~fit for $|b| > 0.5^\circ$ (left) or 2$^\circ$ (right). 
  Solid (dashed) lines correspond to \SFnoGCE~(\SFBBn) diffuse template.  The black line illustrates the  \OppnoGCE\, case. 
  The blue (red) line refers to  \OppB\, (\OppNFW) case. The colored areas correspond to $1\sigma$ uncertainty bands. 
  The black (gray) points represent the count distribution of 4FGL sources
 (without any analysis flag, see \cite{Fermi-LAT:2019yla}).} 
  \label{fig:dNdSall}
\end{figure*}

We then use  \SF-optimized diffuse and smooth GCE templates as input for \Opp~fits, testing 0.5$^\circ$ and 2$^\circ$ latitude cuts. Our results are summarized in Fig.~\ref{fig:dNdSall}, where we show the best-fit \dnds~for the IPS in the IG ROI for several \Opp~fit configurations. 
First, we notice that whatever GCE template is added to the \Opp~fit components (bulge or NFW126), its normalization never converges toward the lower bound of its prior interval, regardless of the \SF~diffuse template adopted. The same is valid for the IPS normalization.  
In all fit setups shown, an IPS population is recovered below the LAT flux threshold. The reconstructed IPS \dnds~is stable against systematics related to the choice of \SF-optimized diffuse template, and latitude cut. %
Moreover, it does not present any spurious effect at the \Fermi-LAT threshold ($\sim 10^{-10}$ ph cm$^{-2}$ s$^{-1}$), and IPS are resolved down to  $\sim 10^{-11}$ ph cm$^{-2}$ s$^{-1}$  for $|b|> 0.5^\circ$, depending on the modeling of the smooth GCE component. 
This holds true even when no GCE smooth template is included neither in the \SF~fit nor in the \Opp~one, contrary to what happens using non-optimized diffuse models~\cite{Leane:2020nmi, Buschmann:2020adf}. 
We therefore demonstrate, also in the context of \Opp~methods, that reducing large-scale residuals from mis-modeling of the diffuse emission improves the reconstruction of PS \dnds~(see also \cite{supplmat}). 
The reconstructed \dnds~has a normalization decreasing with increasing latitude cuts, suggesting that PS are more numerous towards the very GC.
When an  NFW126 template is 
included in the \Opp~fit, the IPS \dnds~is compatible with the \OppnoGCE~case. In both cases, the  second break in the \dnds~-- in addition to the one set in the bright regime -- is recovered close to the LAT flux threshold.
Instead, the \OppB~reconstructs PS down to lower fluxes, regardless of \SFnoGCE~or \SFBBn~diffuse models. For these setups, the second flux break is found at $\sim 2\cdot 10^{-11} \, (8\cdot 10^{-12})$ ph cm$^{-2}$ s$^{-1}$  for $|b|> 0.5^\circ \, (2^\circ)$. Going from $|b|> 0.5^\circ$ to  $|b|> 2^\circ$, the \dnds~is resolved down to even lower fluxes.
A posteriori, we associate such a better sensitivity to IPS to the ability of the fitted diffuse components to further reduce fit residuals.

\begin{table*}
\caption{\textit{Results for the \Opp~analysis of the IG} LAT data. First four columns:  setup of the analysis and  latitude mask of the IG. The $\ln(\mathcal{Z})$ is the nested sampling global log-evidence extracted from \texttt{Multinest} \cite{Zechlin:1}.  Last two columns:  flux percentage of different model components with respect to the total emission in the ROI (for $S<10^{-8}$ \fluxunits, see~\cite{supplmat}), and  normalization of smooth GCE template in the \Opp. Flux percentage always sum to unity within errors.
}

\centering
\begin{tabular}{ l @{\hspace{10px}}| l @{\hspace{10px}}| l @{\hspace{10px}}| c @{\hspace{10px}}|c @{\hspace{10px}} |c  @{\hspace{10px}}| c} \hline\hline

\textbf{Description} &\textbf{\Opp~setup}	& \textbf{\SF~diffuse}	& $|b|$ cut [$^\circ$]& $\ln(\mathcal{Z})$ 	&  Point~sources/diffuse/GCE \%  & $A_{\rm B/NFW126}$ \\ \hline

  No GCE (both) 	&  \OppnoGCE \,	&  \SFnoGCE  & 2	& $-6113$ 	& $12/89/- $ & -\\ 
  Bulge (\Opp~only) & \OppB 	&  \SFnoGCE  & 2	& $-6076 $ 	&  $13/81/7 $ & $0.8\pm0.1$ \\ 
  DM (\Opp~only) &  \OppNFW  	&  \SFnoGCE & 2	& $-6084$ 	&  $10/84/6 $ &  $1.8^{+0.4}_{-0.2}$ \\
                                                             \hline
 Bulge (\SF~only) &  \OppnoGCE \,	&  \SFBBn   & 2	& $ -6169$ 	&  $11/89/- $& - \\  
 Bulge (both) &  \OppB 	&  \SFBBn  & 2	& $-6074$ 	&  $13/77/10 $ & $1.1\pm0.1$ \\ 
DM (both) &  \OppNFW 	& \SFNFWn  & 2	& $-6084$ 	&   $11/82/7 $ & $2.3\pm0.3$ \\ 
 
 \hline\hline
 
 No GCE (both)  &  \OppnoGCE \,	&   \SFnoGCE & 0.5	& $-7822$ 	& $13/86/-$ & - \\ 
Bulge (\Opp~only) &   \OppB 	&  \SFnoGCE  & 0.5	& $-7802$ 	&   $14/83/3$ & $0.3\pm0.1$ \\ 
DM (\Opp~only) &    \OppNFW 	&   \SFnoGCE & 0.5	& $-7818$ 	&   $14/85/1$  & $0.3\pm0.1$ \\ 
    \hline
Bulge (\SF~only) &  \OppnoGCE	&   \SFBBn  & 0.5	& $-7907$ 	&   $15/85/-$  & - \\ 
  Bulge (both) & \OppB 	&   \SFBBn & 0.5	& $-7796$ 	&   $14/79/7$  & $0.8\pm 0.1$ \\ 
  DM (both) &   \OppNFW 	&   \SFNFWn & 0.5	& $-7820$ 	&   $14/84/2$  & $0.6\pm0.2$ \\ 
 \hline\hline
\end{tabular}
\label{tab:results}
\end{table*}
We quantify now the evidence for models with an additional smooth GCE template.
To this end, we compare the global  evidence, $\ln \mathcal{Z}$, for the \OppnoGCE, \OppB~and \OppNFW~setups, with different \SF~diffuse model inputs. For each model combination, 
we compute the Bayes factor between model $i$ and $j$,
$B_{ij} = \exp(\ln \mathcal{Z}_i - \ln \mathcal{Z}_j)$, 
and  
assess the strength of evidence  of model $i$ with respect to model $j$. 
Our results are presented in Tab.~\ref{tab:results}.
Regardless of the \SF-optimized diffuse template adopted, data {\it always} more strongly support models which include an additional smooth template for the bulge with respect to  models without GCE in the \SF~and/or \Opp~fits ($\ln B_{ij}>20$), {\it and} models with an additional smooth NFW126 component in the \SF~and/or \Opp~fits ($\ln B_{ij} > 7$). 
Whenever a bulge template is included in our analysis, this is preferred even with respect to  additional smooth DM templates.
As for our baseline model, \SFBBn, and $|b|>2^\circ$, the evidence for  an additional bulge template (\OppB), with respect to \OppnoGCE~is $\ln B \sim 95$. Moreover, in this case the normalization of the bulge template is $A_{\rm B} = 1.1 \pm 0.1$, supporting the consistency between GCE and diffuse model adopted.
This evidence is as strong also  for $|b|>0.5^\circ$, $\ln B \sim 110$.

\begin{figure*}[t]
  \includegraphics[width=0.49\textwidth]{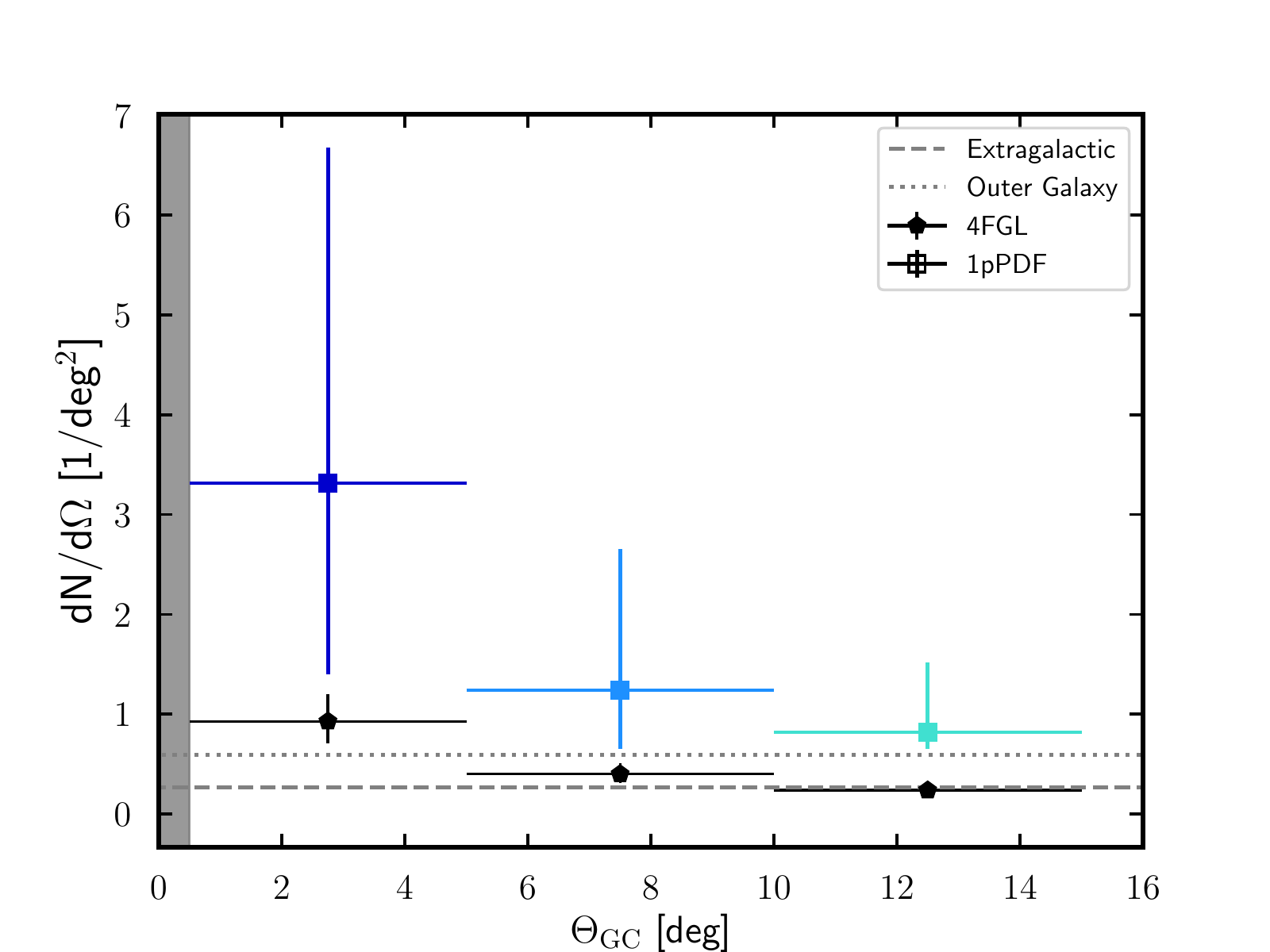}
   \includegraphics[width=0.49\textwidth]{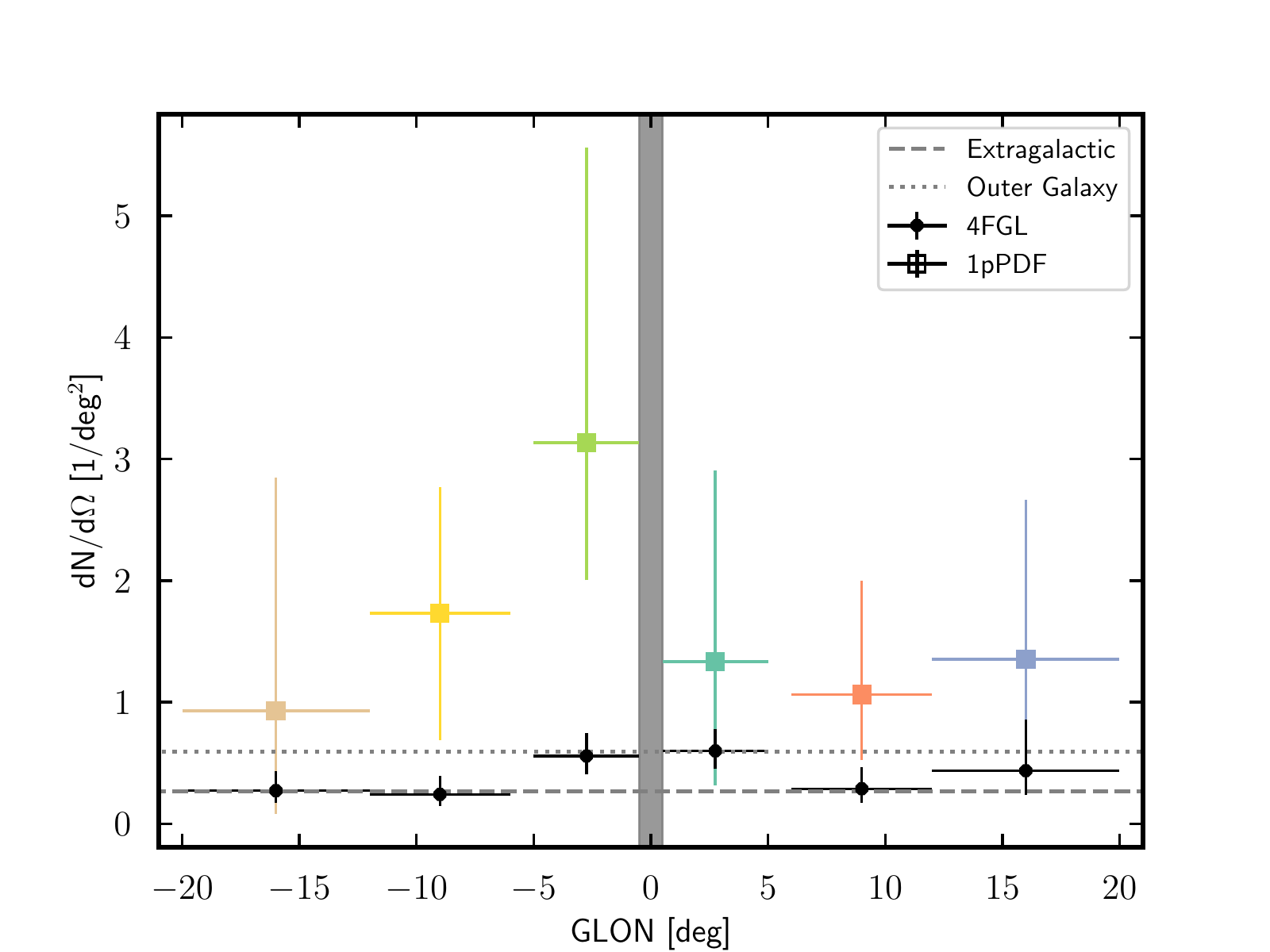}
  \caption{\textit{Radial} (left) \textit{and longitude} (right) \textit{source density $dN/d\Omega$ profiles}, as reconstructed by the \OppB~fit using the \SFBBn~diffuse model. 
  We also display source density profiles for 4FGL sources (black points), and average source densities in the OG and EG ROIs.
  } 
  \label{fig:latprofile}
\end{figure*}

We note that, when we use the \SFnoGCE~diffuse model in the \Opp~fit including the bulge (\OppB), we
find comparable evidence to the \OppB, \SFBBn~setup. 
Indeed,  \SF~is able to re-absorb part of
the photons from the bulge by re-modulating (spatially) other diffuse templates, and so,
partially reduces the residuals also in the \SFnoGCE~case.
This is perfectly consistent with the fact that $A_{\rm B}  = 0.8 \pm 0.1$.
Models with PS {\it and} a smooth bulge component are therefore strongly preferred by data, regardless of the optimized diffuse model employed. 
On the contrary, the evidence for an additional smooth NFW126 template with respect to models without GCE in the \SF~fit and/or \Opp~fits depends on the choice of the \SF-optimized diffuse template adopted, as well as on the latitude cut. 

We have tested our results against a number of systematic effects, which are detailed in \cite{supplmat}. 

The flux percentages reported in Tab.~\ref{tab:results} illustrate that \Opp~fits to \Fermi-LAT data find  non-null (and even comparable) emission from both the IPS population $and$ the smooth GCE template, in most cases each contributing about 10\% of the total emission in the ROI. Since  4FGL sources ($2^\circ$ cut, without analysis flag, see Fig.~\ref{fig:dNdSall}) account for 7\% (10\% including flagged sources) of the total IG emission, the remaining flux  comes from sub-threshold IPS. 
We have verified~\cite{supplmat} that our results are not driven by PS in the ultra-faint regime~\cite{Chang:2019ars}, where the sensitivity of the \Opp~method drops (as quantified by the magnitude of uncertainty bands in Fig.~\ref{fig:dNdSall}), and an IPS population may become degenerate with a truly diffuse  emission.
 
\smallskip

We also
measure the IPS dN/dS in two control regions: The outer Galaxy (OG, $|b|<20^\circ$, $60^\circ<|l|<90^\circ$) and the extragalactic region (EG, $|b|>40^\circ$, $|l|>90^\circ$).
The reconstructed \dnds~in both OG and EG ROIs does not present any spurious threshold effect and can identify IPS down to the statistical limit of the method around $\sim 10^{-11}$ ph cm$^{-2}$ s$^{-1}$ \cite{Zechlin:1}.
 We compute the 
source density $dN/d\Omega$ in the flux interval [$10^{-11}-10^{-9}$] ph cm$^{-2}$ s$^{-1}$, finding $\sim0.6$ sources/deg$^2$ in the OG, and $\sim0.3$ sources/deg$^2$ in the EG, see Fig.~\ref{fig:latprofile}.

Since the spatial distribution of PS is {\it isotropic}  by construction, we test the PS spatial behavior by dissecting the IG ROI into three concentric annuli, masked for latitudes $|b| <0.5^\circ$. 
We extract the \dnds~separately in each ring, and integrate it over the flux interval [$10^{-11} - 10^{-9}$] ph cm$^{-2}$ s$^{-1}$. The result is reported in Fig.~\ref{fig:latprofile} as a function of  the mean $\Theta_{\rm GC}=\sqrt{b^2 +l^2}$ in each ring, for our baseline \OppB, \SFBBn~setup. %
We observe a decreasing trend of the $dN/d\Omega$ in the IG with $\Theta_{\rm GC}$. %
Also, the $dN/d\Omega$ in the innermost ring is about a factor of three higher than 4FGL sources, as well as than in OG and EG. %
For the most external  ring, the source density is instead comparable with the catalog, OG and EG ones. %
This corroborates the evidence that the IG PS population is \textit{not purely isotropic nor extragalactic} in origin, but rather it
peaks towards the GC.
Similarly, we build the longitude profile of IG PS, Fig.~\ref{fig:latprofile}. The \dnds~has been fitted in 6 longitude slices from the GC bound at $|l|=6^\circ, 12^\circ$ and $20^\circ$. The derived $dN/d\Omega$ shows again a distribution peaked around the GC, and compatible with OG (and partially with 4FGL and EG) sources only in the most external longitude interval.
This result adds a piece of evidence that the GCE (defined as an excess of photons above traditionally adopted foreground/background astrophysical models) is contributed by faint PS  on lines-of-sight toward the Galactic center, and, perhaps, in the Galactic bulge, supporting their {\it Galactic} origin.

\medskip

{\it Conclusions.} 
For the first time, we analyzed the IG \Fermi-LAT sky by means of the \Opp~photon-count statistics technique in order to understand the role of PS to the GCE. 
To minimize the systematic effects inherent the modeling of the $\gamma$-ray sky, we introduced important methodological novelties.
First, we implemented within the \Opp~new, \textit{optimized}, models for the diffuse emission from \SF~adaptive template fits, developing a self-consistent procedure which effectively reduces diffuse mis-modeling.
Secondly, besides PS, in the \Opp~fit  we included an additional \textit{smooth} GCE template which traces the stellar distribution in the Galactic bulge.

The updated \SF~analysis of the IG confirms  that the GCE is better described by a bulge template than an NFW126 model at high significance.
Moreover, we find that the \Opp~method, supplied with \SF~diffuse emission templates,  always recovers an IPS population well below the \Fermi-LAT flux threshold, down to  $\sim 10^{-11}$ ph cm$^{-2}$ s$^{-1}$  for $|b|> 0.5^\circ$. 
The reconstructed IPS \dnds~is stable against a number of systematics, in particular related to the choice of \SF-optimized diffuse template and latitude cut. %
Regardless of the \SF-optimized diffuse template, data {\it always} prefer models which include an additional smooth template for the bulge with respect to both models without it and models with an additional NFW126  template, in the \SF~and/or \Opp~fits.

Our results show that, within the statistical validity of the \Opp~and the setups tested, IPS and diffuse bulge each contributes about $\mathcal{O}$(10\%) to the  $\gamma$-ray emission along the lines-of-sight toward the GC.
In particular, within our  baseline model the \Opp~founds that PS (bulge) contribute  13\% (10\%) of the total emission of the  IG. Subtracting  the contribution from cataloged sources, a non-negligible fraction of the IG emission is accounted by sub-threshold PS.  
This further corroborates a possible, at least partial, stellar origin of the GCE.

We also verified  that this IPS population is not purely isotropic nor extragalactic in origin, rather it peaks towards the very GC.
Although the final confirmation of the PS nature of the GCE will most likely come from multiwavelength future observations, we undoubtedly got one step closer to the understanding of the mysterious nature of the GCE emission.

\medskip

\textbf{Acknowledgments.} 
We very kindly acknowledge the work formerly done by H.S. Zechlin on the \Opp~code.
We warmly thank P.~D.~Serpico for inspiring discussion. We also thank M. Di Mauro, F. Kahlhoefer, M. Kraemer, P.~D.~Serpico, and C.~Weniger for a careful reading of the manuscript and for insightful comments.
The work of F.D. has been supported by the ``Departments of Excellence 2018 - 2022" Grant awarded by
the Italian Ministry of Education, University and Research (MIUR) (L.~232/2016).
F.C.~acknowledges support by the Programme National Hautes Energies (PNHE) through the AO INSU 2019, grant ``DMSubG'', and the Agence Nationale de la Recherche AAPG2019, project ``GECO''.
S.M.~acknowledges computing resources granted by RWTH Aachen University under project rwth0578.

\bibliography{biblio}

\begin{thebibliography}{47}
\bibitem [{spa()}]{spatialcomment}%
  \BibitemOpen
  \href@noop {} {\ }\bibinfo {note} {As currently implemented, the \Opp~method
  does not allow to test spatially-dependent \dnds.}\BibitemShut {Stop}%
\end{thebibliography}

\medskip
\onecolumngrid

\setcounter{page}{1}
\renewcommand{\thepage}{S\arabic{page}} 
\renewcommand{\thesection}{S\arabic{section}}  
\renewcommand{\thetable}{S\Roman{table}}  
\renewcommand{\thefigure}{S\arabic{figure}}
\renewcommand{\theequation}{S\arabic{equation}}

\begin{center}

\vspace{0.05in}
{ \large {\it  Supplemental Material:\\
}
\vspace{0.05in}
\bf Dissecting the Inner Galaxy with $\gamma$-Ray Pixel Count Statistics}\\ 
\vspace{0.05in}
{F.Calore, F.Donato, and S.Manconi}
\end{center}

\section{The \SF~analysis}
\label{sm:skyfact}
%
In this section, we discuss in more detail the analysis and results of the \Fermi-LAT $\gamma$-ray fit with \SF~\cite{Storm:2017arh}.

For the \SF~analysis, we consider all \texttt{FRONT+BACK} events (evtype=3) to maximize the statistics against the large number of free parameters in the fit. 
%
The data are binned in energy into 30 logarithmically-spaced bins from 0.2 to 500~GeV, and spatially into cartesian pixels of size 0.5$^\circ$.
The fit is performed in the energy range $0.3-300$~GeV, and the main ROI restricted to the inner 40$^\circ \times 40^ \circ$.
This \SF~ROI is larger than our \Opp~inner Galaxy (IG) ROI since, for the purpose of template fitting, the ROI must be  
large enough to be able to correctly assess the GCE morphology~\cite{Macias:2019omb}.
The model of the $\gamma$-ray sky diffuse components and the statistical analysis closely follow~\cite{Bartels_bulgelum}.
The main novelties here are: (i) The increased data set; (ii) the restricted ROI to ensure stability of the fit with the increased data set; and (iii) the use of the 4FGL catalog to model \Fermi-LAT point-like and extended sources.
%

Every model component is characterized by an \textit{input} spectrum and morphology, which are fitted to $\gamma$-ray data with the adaptive template fitting algorithm implemented in \SF, and based on penalized maximum likelihood regression. 
Hyperparameters in the regularization term of the likelihood control the allowed variation of spectral and spatial free parameters, preventing overfitting. Spectral and spatial modulation parameters (i.e.~nuisance parameters) are allowed to vary to account for mis-modelling of the input templates. The minimization is performed by the L-BFGS-B (Limited memory BFGS with Bound constraints) algorithm. We refer to~\cite{Storm:2017arh} for more details about the technical implementation.
The diffuse $\gamma$-ray (spectral and spatial) model components we input are: (i) An isotropic spatial component with the best-fit IGRB spectrum from~\cite{2015ApJ...799...86A}; (ii) an inverse Compton spectral and spatial component computed for a typical scenario of cosmic-ray sources and propagation parameters with the \texttt{DRAGON} code~\cite{2011ascl.soft06011M}; (iii) three rings for the spatial distribution of photons from $\pi^0$ decay, as traced by the sum of atomic and molecular hydrogen distribution and available within the GALPROP public release\footnote{\url{https://galprop.stanford.edu/}} (the $\pi^0$ decay input spectrum is taken from~\cite{2012ApJ...750....3A}), (iv) the \Fermi~bubbles with spectrum from~\cite{Fermi-LAT:2014sfa} and a uniform geometrical template as input morphology, and (v) the GCE.
For the latter, we test different spatial models: NFW templates with slopes equal to 1 (NFW100) and 1.26 (NFW126) and a bulge template, composed by a boxy-bulge and a nuclear bulge as modeled in~\cite{Bartels_bulgelum}.
Additionally, we refit all point-like and extended sources at the position of 4FGL cataloged sources.
We refer to~\cite{Storm:2017arh, Bartels_bulgelum} for details about the $\gamma$-ray model. 
Spectral and spatial uncertainties on the model components are set by regularization terms. We allow variations as in \texttt{run5} of~\cite{Storm:2017arh} for all components, except for the additional GCE template. For the GCE, we fix the spatial structure of the template (i.e.~no additional freedom allowed on the spatial modulation parameters), while we leave full bin-by-bin freedom to the spectral parameters (i.e.~unconstrained GCE spectrum).
To study systematics on the \dnds, see Sec.~\ref{sm:dNdS_sys},
we run fits for different values of the spatial \textit{smoothing} hyperparameter, $\eta$, for the gas and \Fermi~bubble templates. As defined in~\cite{Storm:2017arh}, the spatial smoothing hyperparameter is $\eta= 1/x^2$, where $x$ is the admitted variation between neighboring pixels. The reference values are $\eta_g=25$ for the gas and $\eta_b=4$ for the \Fermi~bubbles.  By varying the smoothing scale, we therefore check the bias induced by mis-modeling at small scales. 

This updated analysis confirms previous findings about the preference for a bulge morphology of the GCE on top of DM-only templates. In Tab.~\ref{tab:SFresults}, we report the log-likelihood values of \SF~fits with different GCE templates. 
From the likelihood values, using the $\delta-\chi^2$ statistics~\cite{Bartels_bulgelum}, one can compute the evidence for any additional template in nested models. In this case, we find that: Adding an NFW template (no matter the slope) on top of a model with the bulge already included does not significantly improve the fit ($3.3\sigma$ for NFW100, and $4.1\sigma$ for NFW126), while adding a bulge template on top of an NFW-only model does ($12.1\sigma$ for NFW100, and $9.8\sigma$ for NFW126).
In general, NFW100 provides larger residuals than NFW126.

Finally, we also run \SF~fits in the OG control region ($|b|<20^\circ$, $60^\circ<|l|<90^\circ$), where we do not find evidence for the additional bulge component, see Tab.~\ref{tab:SFresults}.
%
\begin{table}[h]
    \centering
        \caption{\textit{Log-likelihood values for \SF~fits} with various GCE templates.  Results for the 40$^\circ \times 40^\circ$ IG ROI and for the OG ROI, for an unconstrained GCE spectrum.  }
    \begin{tabular}{c|c|c}
      \toprule
      ROI & \SF~run & $-2\ln\mathcal{L}$ \\
      \hline
      IG &	\texttt{r5\_noGCE} & 151770.6 \\
     &	\texttt{r5\_NFW126} & 151616.6 \\
	& \texttt{r5\_NFW100} & 151686.6 \\
	& \texttt{r5\_bulge} & 151482.8 \\
	& \texttt{r5\_bulge\_NFW126} & 151435.4\\
	& \texttt{r5\_bulge\_NFW100} & 151445.2 \\
      \hline
      OG &	\texttt{r5\_noGCE} &  210753.6\\
	& \texttt{r5\_bulge} & 210740.9 \\
      \hline
      \hline
      \end{tabular}
\label{tab:SFresults}
\end{table}

\section{The \Opp~analysis}\label{sm:1ppdf}
In this section, we provide further details on the \Opp~analysis. 
We start by a small review of the \Opp~method and implementation as presented in ~\cite{Zechlin:1,Zechlin:2,Zechlin:3}, with some details specifically important for the analysis of the IG. 
The \Fermi-LAT dataset used for the \Opp~analysis is then described, before focusing on the model parameters and priors for the \Opp~analysis. 

\subsection{The \Opp~method}
Different implementations of photon-count statistical methods applied on \Fermi-LAT data have been presented \cite{Zechlin:1} (also called \Opp), and \cite{Mishra-Sharma:2016gis} (also called NPTF). 
The \Opp~method have been extensively presented in ~\cite{Zechlin:1,Zechlin:2,Zechlin:3}, to which we refer for any further detail. 
We here briefly highlight the variations introduced in this work for the application to the IG with \SF~templates.

In the actual implementation of the \Opp, the probability $p_k^{(p)}$ of finding $k$ photons  is pixel-dependent, $p$ denoting the evaluated map pixel. 
Different photon sources will contribute to the $p_k^{(p)}$ with different statistics.
Truly diffuse, isotropic emissions will contribute to $p_k^{(p)}$ with counts following a Poissonian distribution. 
The presence of non-Poissonian sources, such as PS, and more complex diffuse structures alters the shape of $p_k$, which permits to investigate these components by means of the \Opp~of the data (see ~\cite{Zechlin:1} for details).

The total diffuse contribution is given by the sum of the Galactic diffuse emission (as derived by \SF~or taken from existing models, see next), a diffuse component describing the GCE as derived by \SF~following a bulge or DM morphology, and a truly isotropic diffuse emission.
 Being  $x^{(p)}_\mathrm{diff}$ the number of diffuse photon
counts expected in a map pixel $p$ we have: (see also \cite{Zechlin:3})
\begin{equation}\label{eq:xdiff}
  x_\mathrm{diff}^{(p)} = A_\mathrm{gal} x_\mathrm{gal}^{(p)} +
  A_\mathrm{GCE} x_\mathrm{GCE}^{(p)} + \frac{x_\mathrm{iso}^{(p)}}{F_\mathrm{iso}} F_\mathrm{iso}.
\end{equation}
$F_\mathrm{iso}$ is the integral flux of the isotropic diffuse emission, and the expression in Eq.\eqref{eq:xdiff} permits to use directly $F_\mathrm{iso}$ as a sampling parameter, in order to have physical units of flux. 
The \SF~templates for the Galactic diffuse emission and GCE 
enter  the \Opp~fit with the best-fit normalization as found within the corresponding \SF~analysis, and we allow  $A_\mathrm{gal, GCE}$ as additional normalization factors.
When dealing with real $\gamma$-ray data, one has to take into account
source-smearing effects  due to a finite detector point-spread
function (PSF), which cause  the photon flux detected from a given point source 
to be spread over a certain area of the sky (i.e. adjacent pixels, when skymaps are divided in pixels). We correct for the PSF effect by statistical means as detailed in \cite{Zechlin:1}. 

\subsection{\Fermi-LAT dataset for \Opp}
\begin{figure*}[t]
  \includegraphics[width=0.32\textwidth]{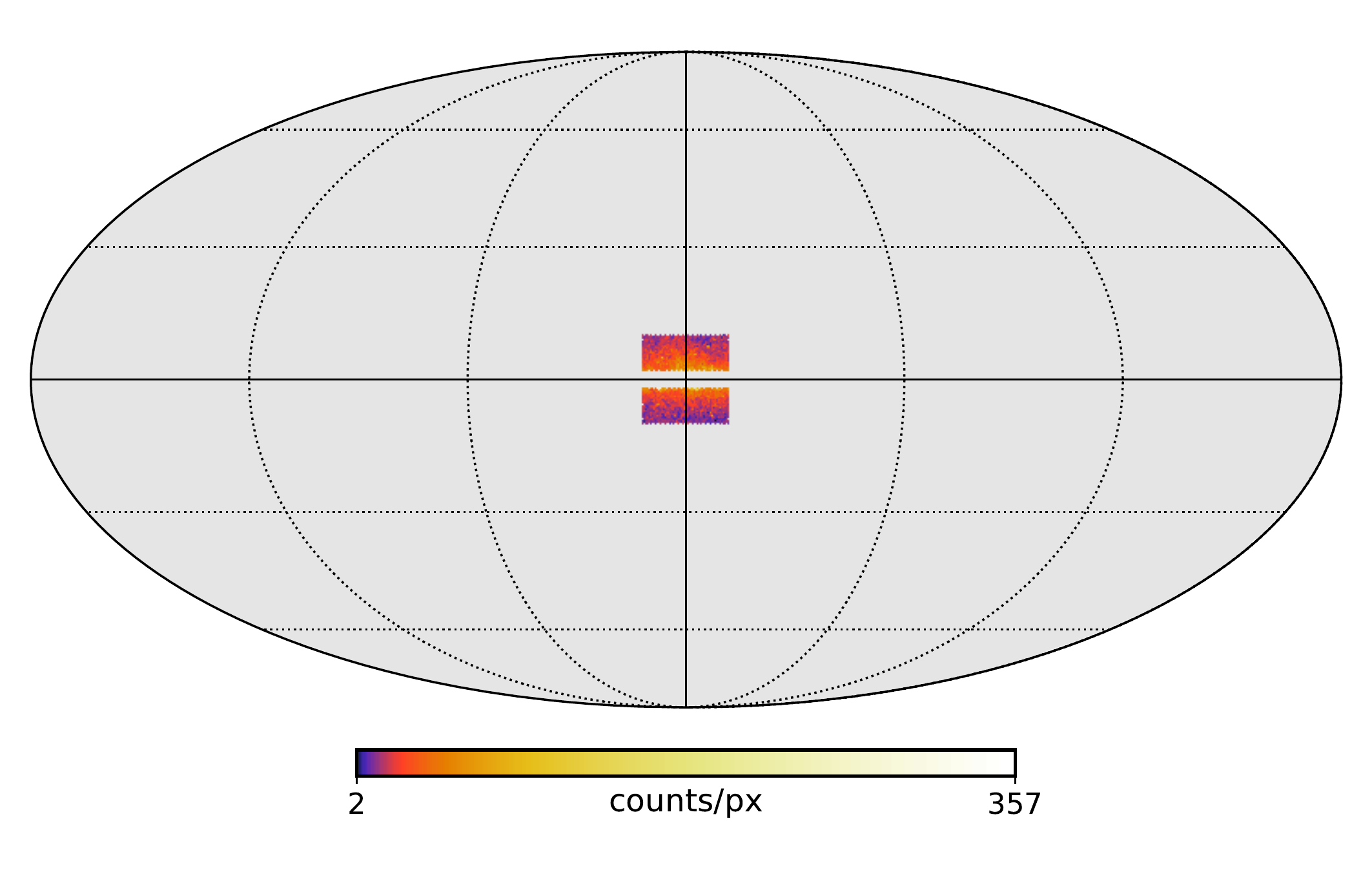}
    \includegraphics[width=0.32\textwidth]{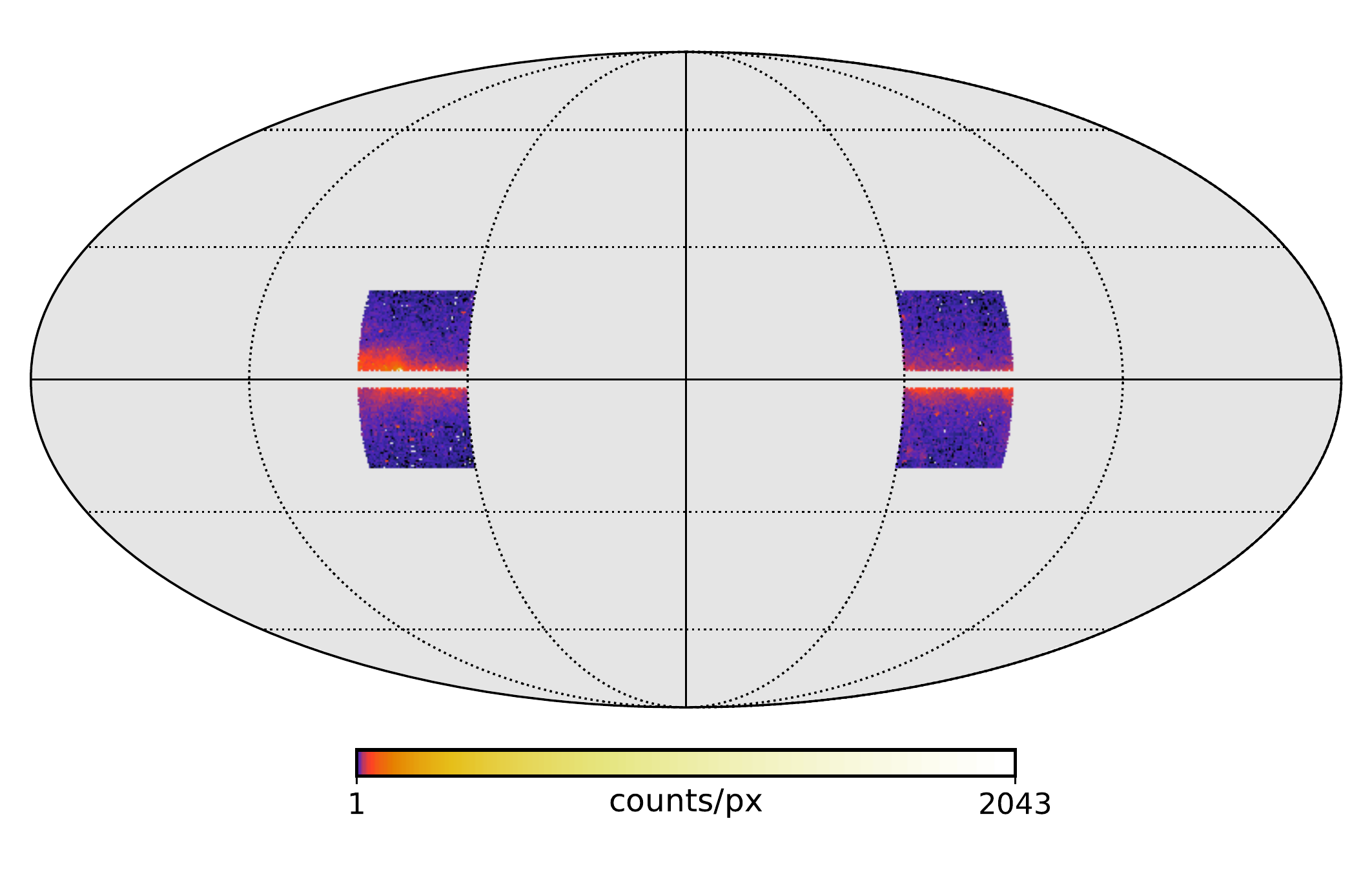}
     \includegraphics[width=0.32\textwidth]{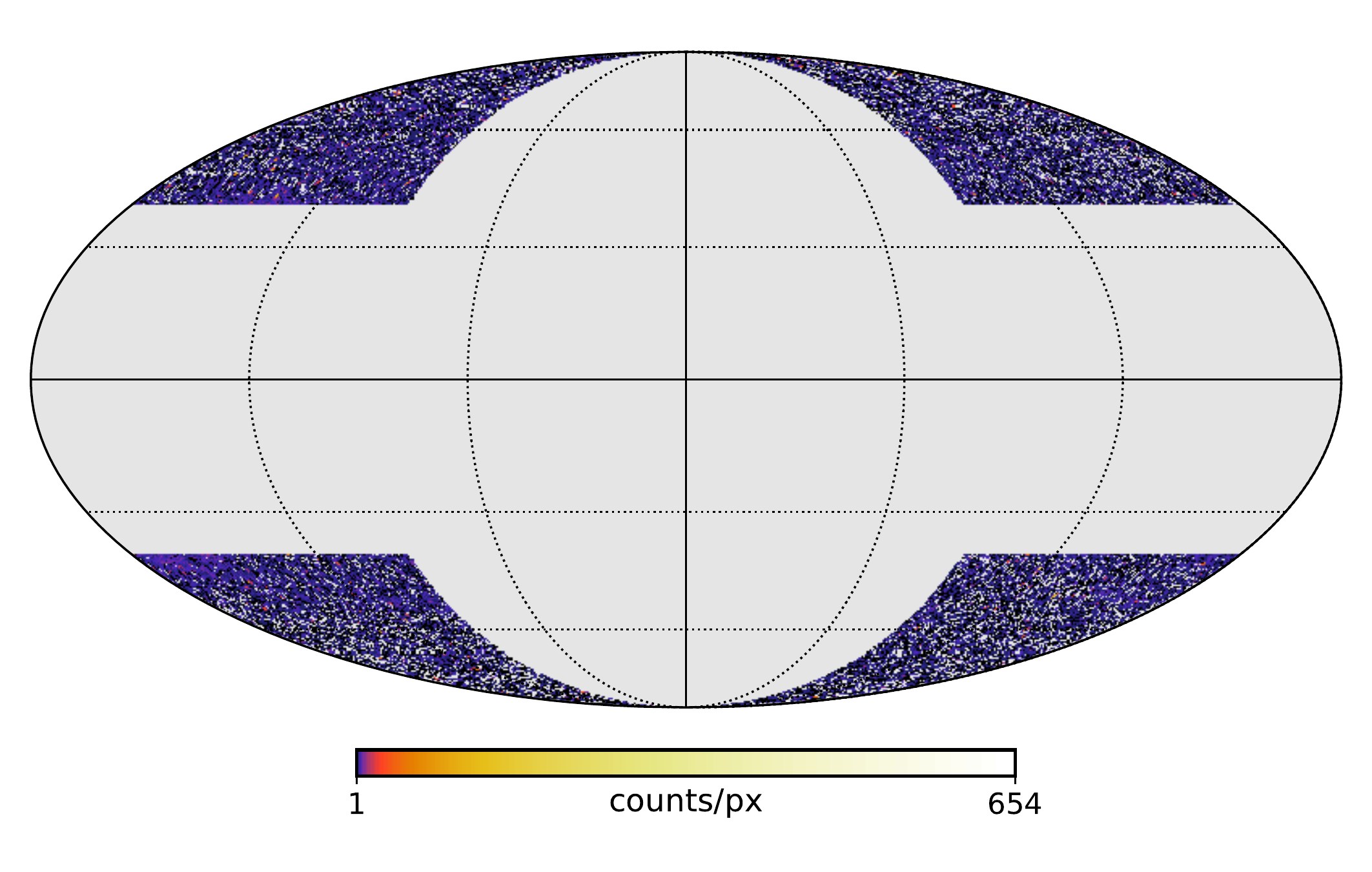}
  \caption{Left panel: Region of interest for the \Opp \,analysis in the \textit{Inner Galaxy}. Middle panel: \textit{Outer Galaxy}. Right panel: \textit{extragalactic}. The \Fermi-LAT counts per pixel in the energy bin [2, 5] GeV are reported.}
  \label{fig:ROIs}
\end{figure*}
We  restrict the analysis of \Fermi-LAT data to events in the quartile with the best angular reconstruction, i.e. evtype=PSF3. 
The spatial binning of photon events is performed using the HEALPix equal-area pixelation scheme \cite{2005ApJ...622..759G} with resoluton parameter $\kappa=7$ \cite{2005ApJ...622..759G} (the total number of pixels covering the entire sky is $N_{\rm pix}=12 N^2_{\rm side}$ and $N_{\rm side}=2^\kappa$).
The analysis is restricted to the $2-5$ GeV energy bin. Previous analyses of the IG using photon-count statistics typically started from 2 GeV to avoid significant PSF smoothing and Galactic diffuse emission systematics at lower energies, while going up to $20$~GeV \cite{Lee:2015fea,Buschmann:2020adf}.
We choose to cut at $5$ GeV instead of $20$~GeV as the photon statistics is largely dominated by lower energies, and to easily compare to our previous results at high Galactic latitudes \cite{Zechlin:1,Zechlin:2}. 
Also, performing the \Opp \,analysis in smaller energy bins might allow to further investigate the nature of IPS, which is left to future work. 

A benchmark upper flux cut of $10^{-8}$ ph cm$^{-2}$ s$^{-1}$ is applied. 
We do not mask resolved sources, but we model them together with fainter PS to reduce possible systematics connected to the source masking or to additional Poissonian templates. 
A lower flux cut is only effectively applied  within the hybrid method, see next section.

\begin{figure*}[t]
  \includegraphics[width=0.48\textwidth]{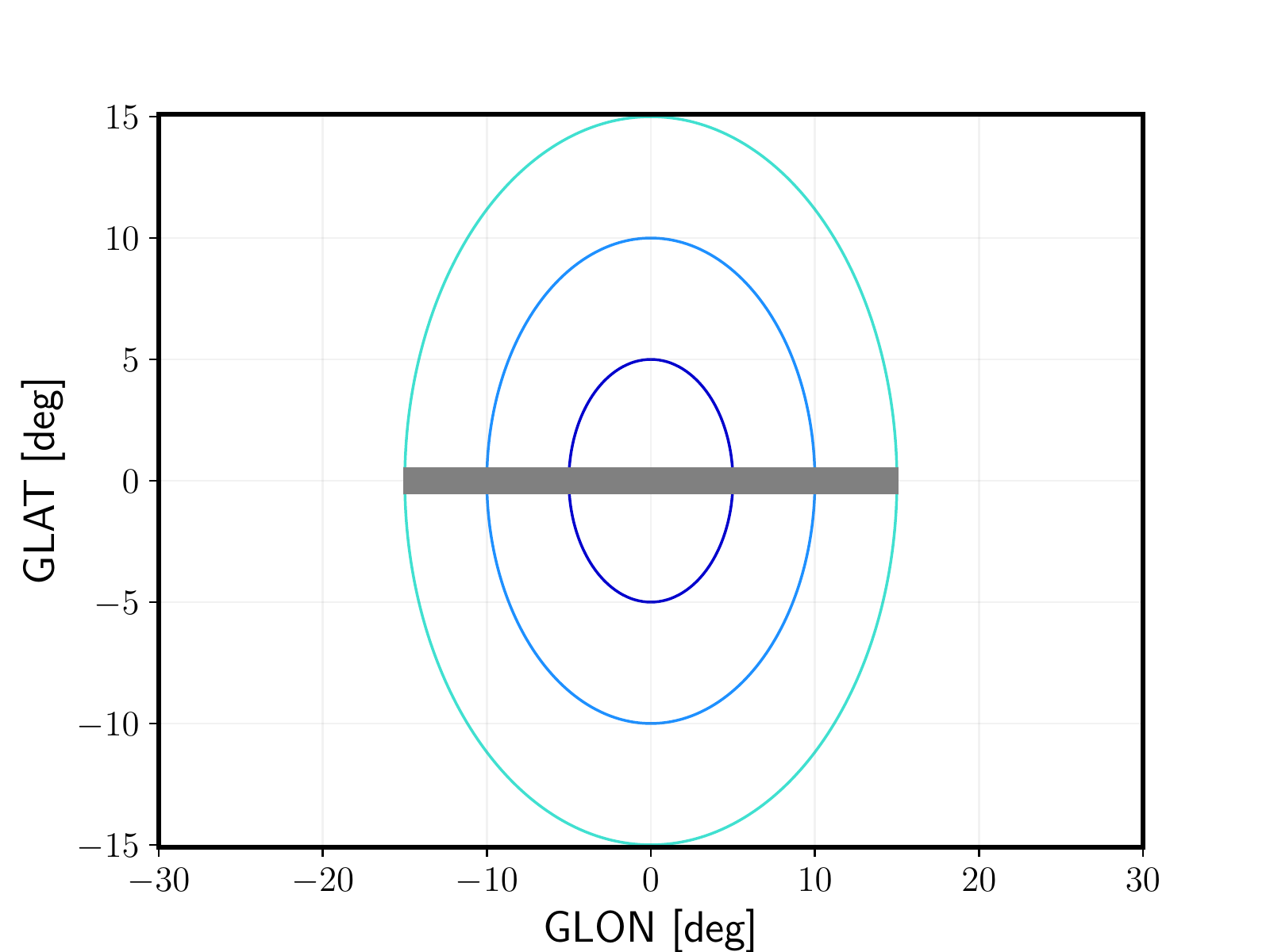}
 \includegraphics[width=0.48\textwidth]{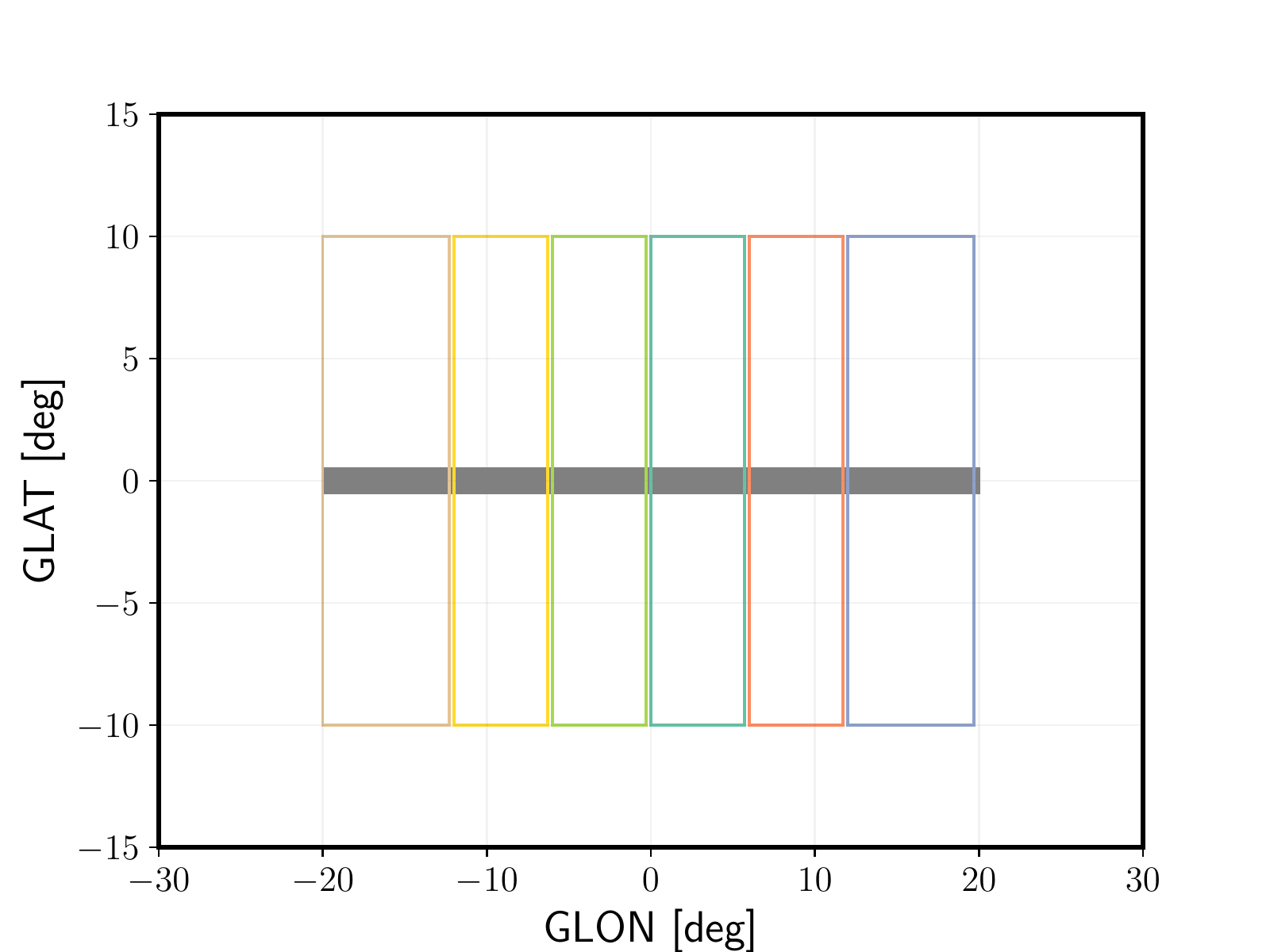}
  \caption{\textit{Radial and longitude slices.}
 The Inner Galaxy division in radial (longitude) slices in the left (right) panel used for computing the source number density depicted in Fig.~2, with consistent color coding.} 
  \label{fig:slices}
\end{figure*}

To dissect the source-count distribution of PS in the IG we restrict to specific ROI. 
The fiducial IG ROI is defined by a $20^\circ \times 20^\circ$ square around the GC. As discussed in \cite{Buschmann:2020adf}, the choice of the ROI can influence the results of photon-count statistical analysis, and needs to be chosen carefully. 
We found the $20^\circ \times 20^\circ$ square optimal to contain enough statistics to constrain the IPS parameters, and not too large, to avoid possible over-subtraction of the Galactic diffuse emission \cite{Buschmann:2020adf}. 
Nevertheless, we tested that our main conclusions are unchanged when going to $30^\circ \times 30^\circ$. 
Although we cannot test directly in the fit the preference for different PS spatial distributions, we can nevertheless verify if these PS are truly isotropic over the ten-degree scale of the IG ROI, and if they are mainly Galactic or extragalactic. To this end,  the
 IG ROI is partitioned in radial and longitudinal slices to build the profiles shown in Fig.~2.
For the sake of definiteness, the corresponding portions of the IG are illustrated in Fig.~\ref{fig:slices}.

Furthermore, two additional ROIs are used to investigate the nature of the IPS found in the IG ROI: The outer Galaxy (OG, $|b|<20^\circ$, $60^\circ<|l|<90^\circ$) and the extragalactic ((EG, $|b|>40^\circ$, $|l|>90^\circ$) ROIs.
The first is meant to represent Galactic IPS, but away from the GC, while the second to provide the truly extragalactic population of IPS.  
We define the OG such that, from the \SF~runs, we do not have any longer evidence for a GCE emission.
Since there is no evidence for the GCE,  the diffuse models obtained with and without this additional component are consistent. 
For the IG and OG ROIs we also cut the innermost region along the Galactic plane. We test different cuts $|b|>0.5^\circ, 2^\circ, 4^\circ$, finding consistent results.  
We stress that a latitude cut of $2^\circ$ was used in past analyses \cite{Lee:2015fea,Buschmann:2020adf,Leane:2020nmi}, while we demonstrate the consistency of our results down to $0.5^\circ$ for the first time. 

An illustration of these three ROIs is provided in Fig.~\ref{fig:ROIs}. 
For each ROI, the counts per pixel of \Fermi-LAT data in the energy bin $2-5$~GeV and for the analysis cuts described in this section are reported. 

\subsection{Model parameters, priors, fitting procedure}
We recall that the \Opp~fits to \Fermi-LAT data are performed with the following components:  
Isotropic diffuse emission template, 
diffuse emission template (optimized or not through \SF~fit), IPS population with source-count distribution per unit flux \dnds~and, if included, a diffuse template describing the GCE, following a Galactic bulge or DM morphology. 
The IPS population is described by a unique multiple broken power law (MBPL, see Eq.~(1)) with two or three free breaks, while the Poissonian components have one free normalization each.
The normalization constant in Eq.~(1) is fixed to $S_0=5\cdot 10^{-9}$ \fluxunits. 
The results presented in the main text have been obtained using a Galactic diffusion emission template optimized by \SF, to reduce background systematics. Results using alternative diffuse emission templates are discussed in the following. 
Both the diffuse emission templates and the GCE templates used as inputs for the  \Opp~fits are the optimized outputs of \SF, and enter the \Opp~fits with an additional free normalization,  $A_{\rm gal}$ and $A_{\rm B/NFW126}$, respectively.
The pixel-dependent likelihood function $\mathcal{L}({\bf \Theta})$ is defined following the L2 method in \cite{Zechlin:1}.
In this way, the spatial morphology of the \SF~diffuse templates is taken into account in the \Opp~fits. 
In fact,  while below the sensitivity of the \Opp~method the PS in the ultra-faint regime are in principle degenerate with a diffuse emission, our pixel dependent likelihood is sensitive to the morphology of the diffuse spatial templates.
The full list of free parameters $\Theta$, along with their prior intervals  are summarized in Tab.  \ref{tab:priors}. 

\begin{table}
\caption{ Prior types and ranges for the  \Opp~analysis of the IG. The first two parameters are in common with all the setups, while the third is present only in the \OppB, \OppNFW~analyses. 
The second block refers to the parameters for the IPS when using the MBPL fit approach, while the last one to the Hybrid approach, where the MBPL was extended with a node. The normalizations $A_S, \, A_{\rm nd1}$ are given in units of  s~cm$^{2}$sr$^{-1}$. The break positions $S_{\rm bn, snd1}$ are in units of \fluxunits. The $F_{\rm iso}$ is given in units of cm$^{2}$s$^{-1}$sr$^{-1}$. All other parameters are dimensionless. }
\centering
\begin{tabular}{ l @{\hspace{10px}}| l @{\hspace{10px}}| l @{\hspace{10px}}| c} \hline\hline
Method	& Parameter	& Prior& Range $N_b=2$ (=3)  \\ \hline
  & $A_{\rm gal}$	& log-flat & [0.1,10]\\ 
    	& $F_{\rm iso}$	& log-flat & [$10^{-11}$, $10^{-8}$]\\ 
    		& $A_{\rm B/NFW126}$& log-flat & [$10^{-2}$, $10^{2}$]\\ \hline
    	
 MBPL   & $A_S$	& log-flat & [$10^{8}$, $5\cdot10^{11}$]\\ 
     	& $S_{\rm b1}$	& log-flat & [$5\cdot10^{-9}$, $10^{-8}$]\\ 
     	& $S_{\rm b2}$	& log-flat & [$10^{-13}$ ($5\cdot10^{-10}$), $5\cdot10^{-9}$] \\ 
     	& $S_{\rm b3}$	& log-flat &  [\dots($10^{-13}$), \dots ($5\cdot10^{-10}$)]\\ 
     	& $n_{\rm 1}$	& flat & [-1, 3]\\ 
     	& $n_{\rm 2}$	& flat & [-1, 3.5]\\ 
     	& $n_{\rm 3}$	& flat & [-2 (1), 2 (3)] \\ 
     	& $n_{\rm 4}$	& flat &  [\dots (-2),\dots (2)]\\ \hline
Hybrid  & $A_S$	& log-flat & [$10^{8}$, $5\cdot10^{11}$]\\ 
     	& $S_{\rm b1}$	& log-flat & [$10^{-9}$, $10^{-8}$]\\ 
     	& $S_{\rm b2}$	& log-flat & [$10^{-11}$($2\cdot10^{-10}$), $10^{-9}$]\\ 
     	& $S_{\rm b3}$	& log-flat & [\dots ($10^{-11}$), \dots($2\cdot10^{-10}$)]\\ 
     	& $n_{\rm 1}$	& flat  & [2.5 ,4.3]\\ 
     	& $n_{\rm 2}$	& flat  & [1.3,2.3]\\ 
     	& $n_{\rm 3}$	& flat  & [1.3,3]\\ 
     	& $n_{\rm 4}$	& flat  & [\dots (1.3),\dots (3)]\\ 
     	& $A_{\rm nd1}$	& log-flat & [$10^{14}$, $10^{17}$]\\ 
     	& $S_{\rm nd1}$	& fixed & $5\cdot10^{-12}$($3\cdot10^{-12}$)\\ 
     	& $n_{\rm f}$	& fixed & -10\\ \hline\hline
\end{tabular}
\label{tab:priors}
\end{table}

To sample the posterior distribution $P({\bf \Theta}) = \mathcal{L}({\bf
  \Theta}) \pi({\bf \Theta}) / \mathcal{Z}$ (where $\pi({\bf \Theta})$
is the prior and $\mathcal{Z}$ is the Bayesian
evidence) the \texttt{MultiNest} framework \cite{2009MNRAS.398.1601F} was used in its standard configuration, setting 1000 live points with a tolerance criterion of 0.2. 
One-dimensional profile
likelihood functions \cite{2005NIMPA.551..493R} for each parameter are built from the final posterior sample in order to get prior-independent frequentist parameter
estimates. If not differently stated, best-fit parameter estimates refer to the obtained maximum likelihood parameter values. Consistent results are found for the Bayesian parameter estimation.
The nested sampling global log-evidence  $\ln(\mathcal{Z})$ is used to build Bayes factors for model comparison, see main text. 

Different fitting techniques have been introduced within the \Opp~framework \cite{Zechlin:1}. We here use the MBPL approach as benchmark, where the parameters of the MBPL in Eq.~(1) are sampled directly. 
To check for possible systematics introduced in the ultra-faint regime, we also perform our  analysis using the Hybrid approach (see \cite{Zechlin:1} for details).
This is characterized by fixing a grid number
of nodes (i.e.~fixed flux positions) for the \dnds~fit, around the sensitivity threshold of the analysis. 
The Hybrid approach was introduced to address possible underestimation or bias in the reconstructed source-count distribution fit and its uncertainty bands at the lower end of the faint-source regime, as demonstrated using Montecarlo simulations in \cite{Zechlin:1}. 
This  has been  shown  to alleviate possible bias when measuring the \dnds~in the ultra-faint regime, which can cause a loss of sensitivity of the method.

\section{Diffuse emission template systematics}
\label{sm:diffuse_sys}
\begin{figure*}[t]
  \includegraphics[width=0.48\textwidth]{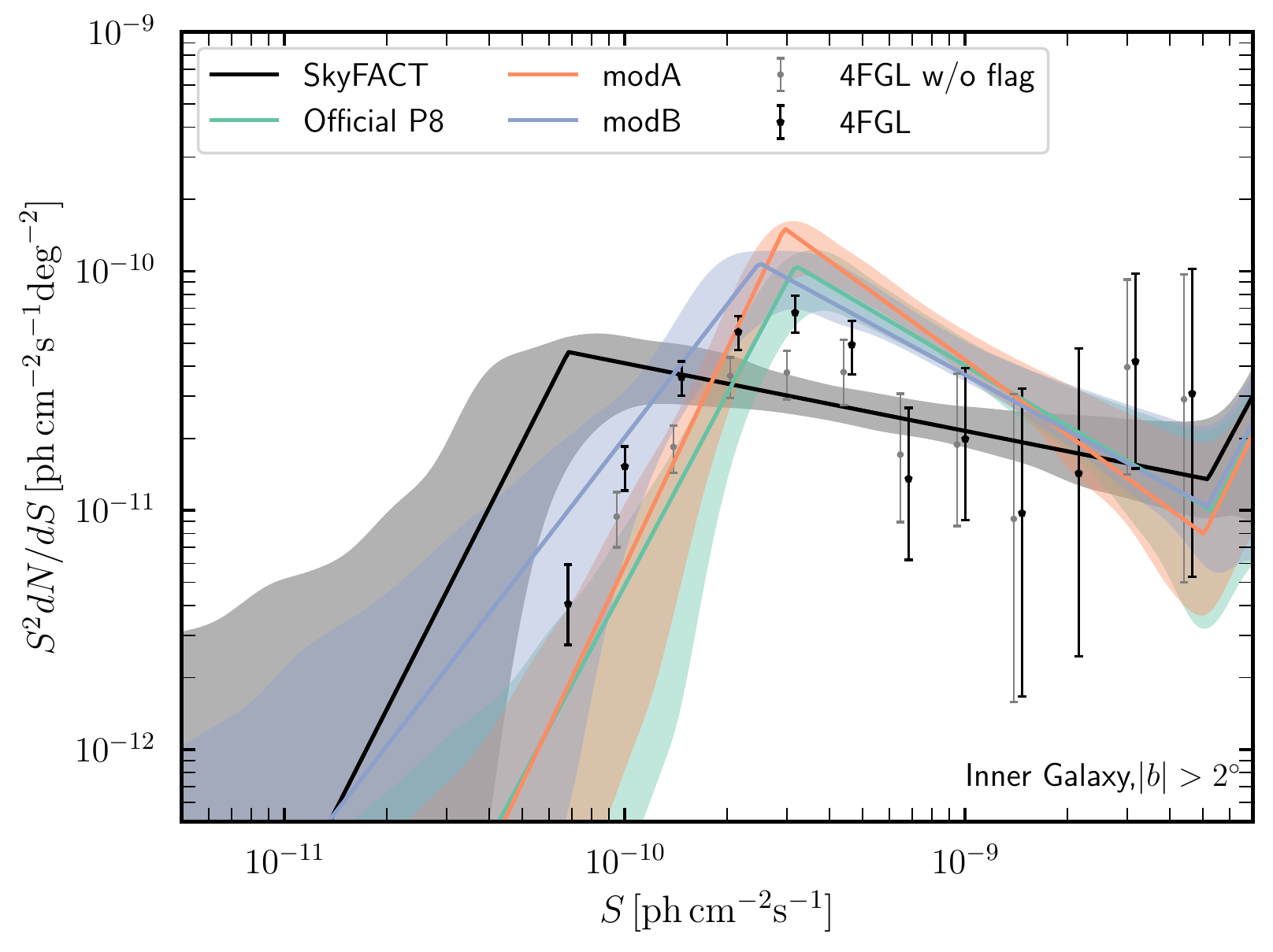}
  \includegraphics[width=0.48\textwidth]{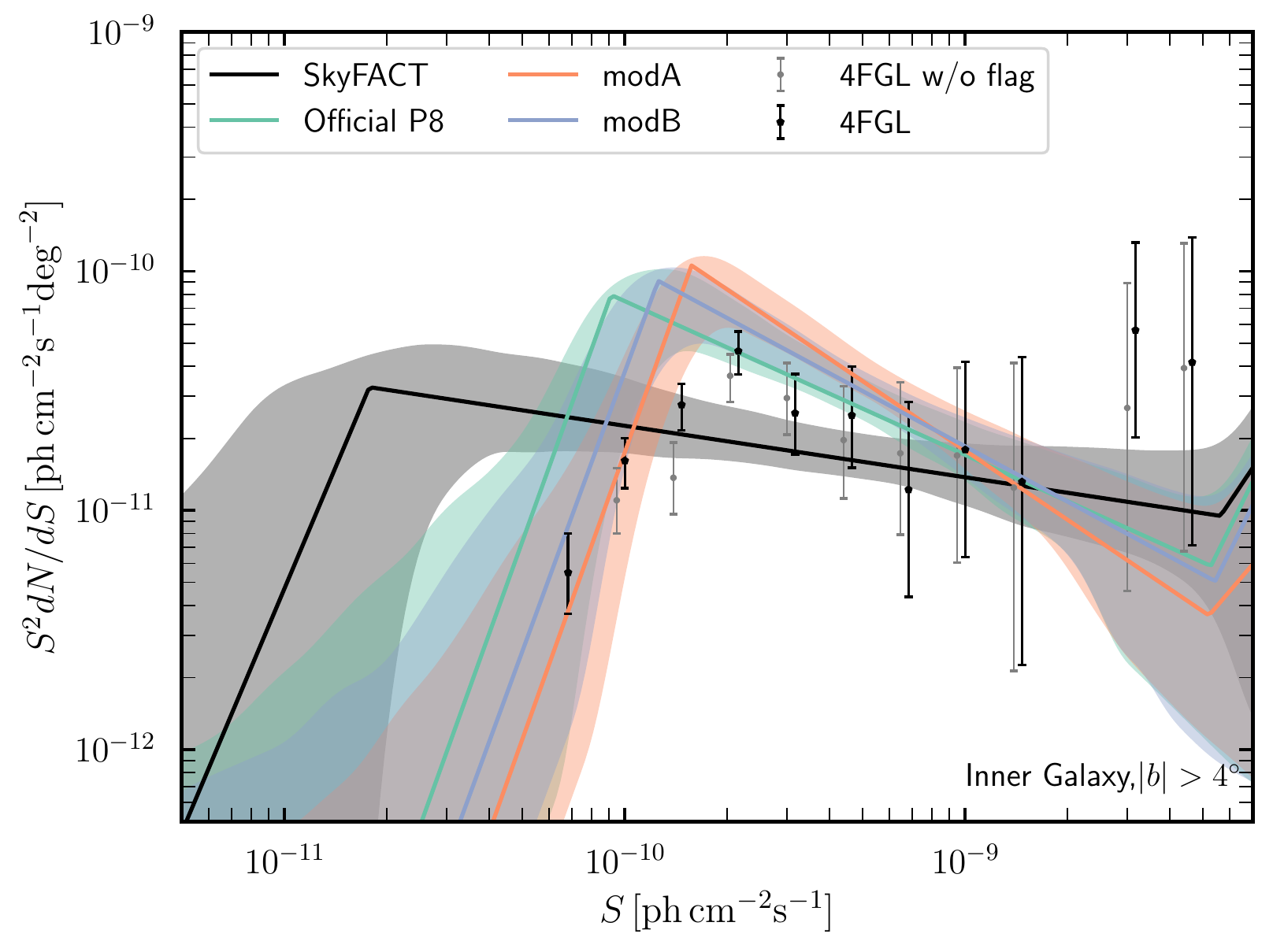}
  \caption{\textit{Diffuse emission systematics.} Source count distribution in the IG obtained from the \Opp~analysis cutting the inner $2^\circ$ (left panel) and $4^\circ$(right panel). 
  The black line is obtained from the \Opp~when using the model for the Galactic diffuse emission obtained from \SF~(without any component modeling the GCE, \SFnoGCE). The colored lines are instead obtained from the \Opp~using the official \Fermi-LAT model for \texttt{Pass 8} (cyan line), or modA and modB (orange and indaco lines). 
 The black (gray) points represent the count distribution of 4FGL sources
 (without any analysis flag, intended as a cautionary index for the reality of a source or the magnitude of its systematic uncertainties \cite{Fermi-LAT:2019yla}).}
  \label{fig:GDEmodel_syst}
\end{figure*}

One of the main novelties of this work is the fact to employ consistently diffuse emission models optimized using \SF~within the \Opp. 
We here apply the \Opp~to the IG using other widely used diffuse emission templates: The official spatial and spectral template released by the \textit{Fermi}-LAT Collaboration for \texttt{Pass 8} data (Official P8) ({\tt gll\_iem\_v06.fits}, see
Ref.~\cite{2016ApJS..223...26A}),
and the models labeled A (modA) and B (modB), optimized for the study of the IGRB in~\cite{2015ApJ...799...86A}. 
We note that within  modA and modB  the \Fermi~bubbles are not modeled.
The results for the \dnds~of the IG are illustrated in the left (right) panel of Fig.~\ref{fig:GDEmodel_syst} when cutting the innermost $2^\circ$ ($4^\circ$). 

By using standard diffuse models (modA, modB and Official P8), we reconstruct spurious sources at $\sim 4 \times 10^{-10}$ ph cm$^{-2}$ s$^{-1}$, well above the sensitivity of the \Opp. 
Such a peak of the IPS \dnds~disappears instead if we use diffuse emission templates as optimized with \SF. 
Large scale residuals are indeed reduced when allowing the spatial diffuse templates to be remodulated in the fit. Even in the absence of an additional GCE template, the \SF~fit remodulates the diffuse components such to partially absorbs GCE photons, therefore reducing residuals and improving the fit with respect to standard diffuse models.
Also, in this ROI, all the diffuse models, except the \SF~one, do not properly reproduce the 4FGL catalog bright sources.  
We notice that the spurious IPS peak of the \dnds~corresponds to a peak of flagged 4FGL sources, further corroborating the conclusion that it is indeed a spurious reconstruction effect. 
We stress that any comparison with 4FGL cataloged sources is purely illustrative, and serves for cross-checking our results at high fluxes.

We therefore confirm previous findings \cite{Buschmann:2020adf} that large residuals due to mis-modelling of diffuse emission 
induce a bias in the reconstruction of PS in the inner Galaxy. 

\begin{figure*}[t]
  \includegraphics[width=0.48\textwidth]{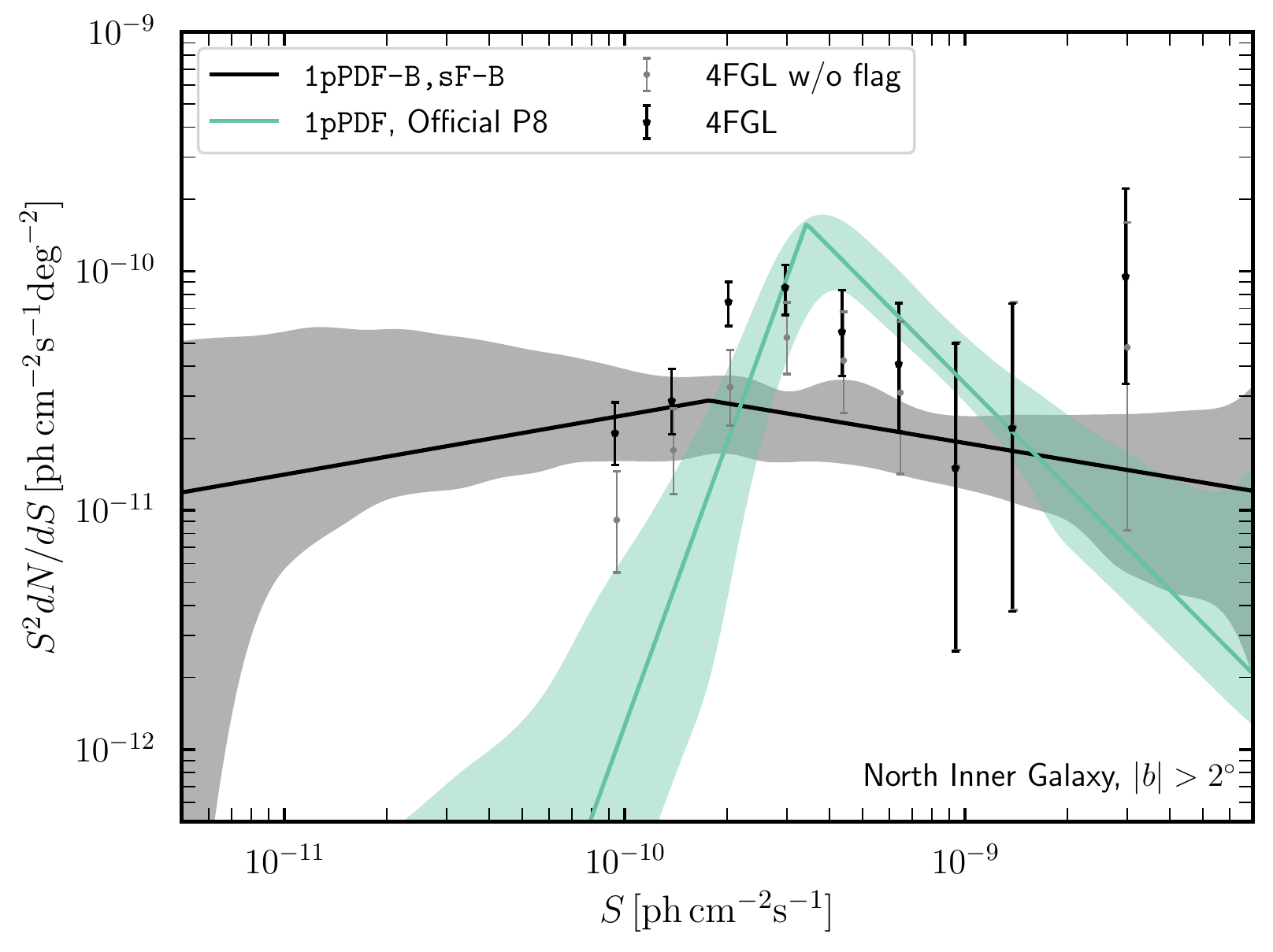}
    \includegraphics[width=0.48\textwidth]{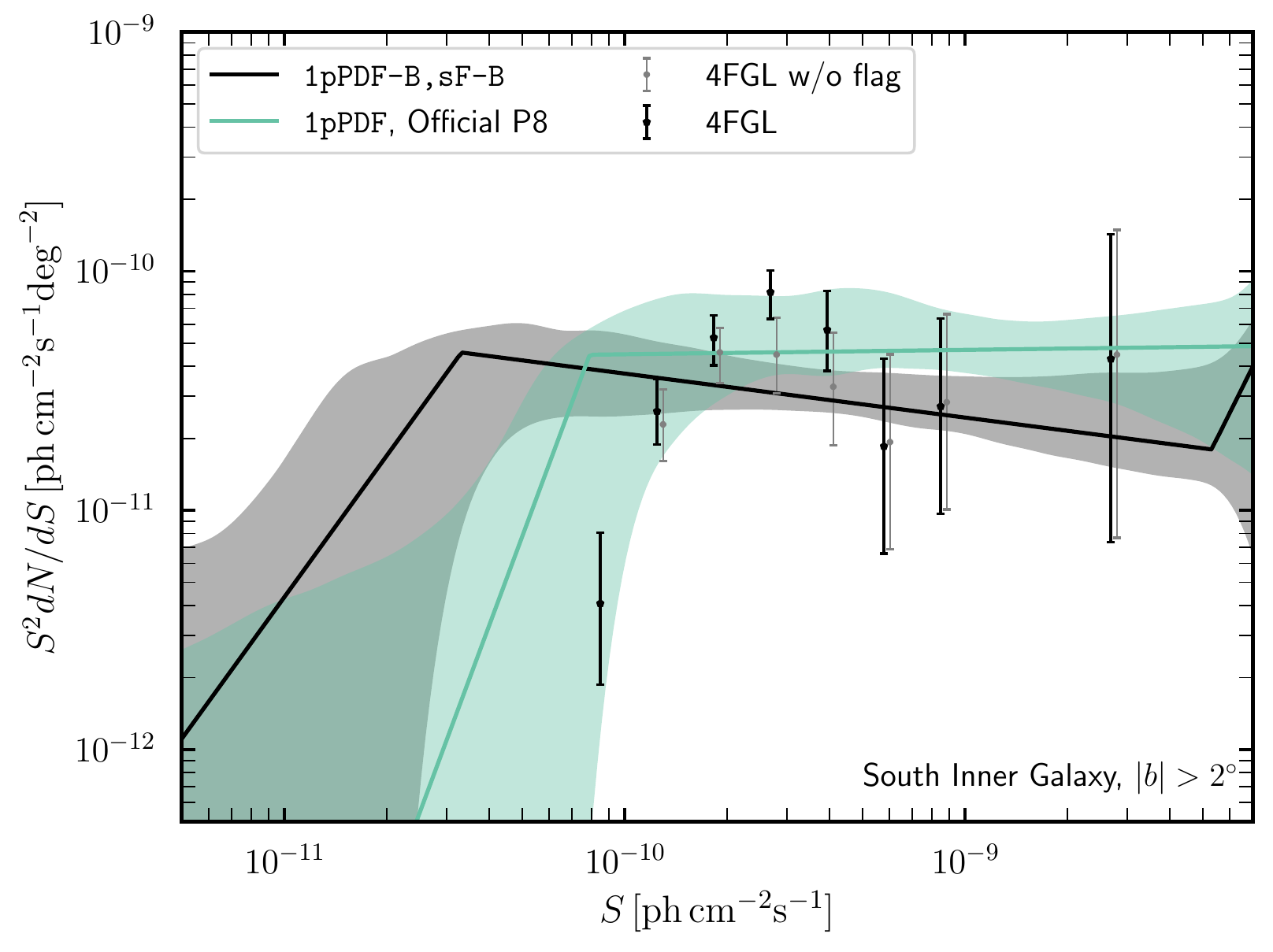}
\caption{\textit{North and South Inner Galaxy.} Left (right) panel: Source-count distribution of the North (South) region of the IG obtained from the \Opp~ analysis. 
Results are here reported using the Official P8 and the \SFBBn~models for the diffuse emission. 
Points as in Fig.~\ref{fig:GDEmodel_syst}.}
  \label{fig:NS}
\end{figure*}

We also identify spatially critical regions within the IG where this mis-modeling effect is more pronounced, notably the Northern hemisphere (both West and East quadrants).
This might be connected to the North/South asymmetry found within the NPTF analysis of the GCE discussed in \cite{Leane:2020nmi}.
As shown in Fig.~\ref{fig:NS} the spurious IPS peak of the \dnds~reconstructed with the \Opp~using the Official P8 template is found to be strongly pronounced in the North IG ROI in the same flux region as found in Fig.~\ref{fig:GDEmodel_syst}, while it is not present in the analysis of the South IG ROI.  
When using the diffuse emission templates as obtained with \SF, we find instead a smoother \dnds~which is compatible within $1\sigma$ uncertainty with the 4FGL unflagged sources.
Moreover, the \Opp~results for the \dnds~using the Official P8 and the \SF~diffuse templates in the South IG ROI are  compatible within the obtained $1\sigma$ bands.

\section{\dnds~modeling systematics}
\label{sm:dNdS_sys}
The stability of the \dnds~results in the IG from the combined \Opp-\SF~analysis of \textit{Fermi}-LAT data was tested against a number of systematics. 
\begin{figure*}[t]
  \includegraphics[width=0.48\textwidth]{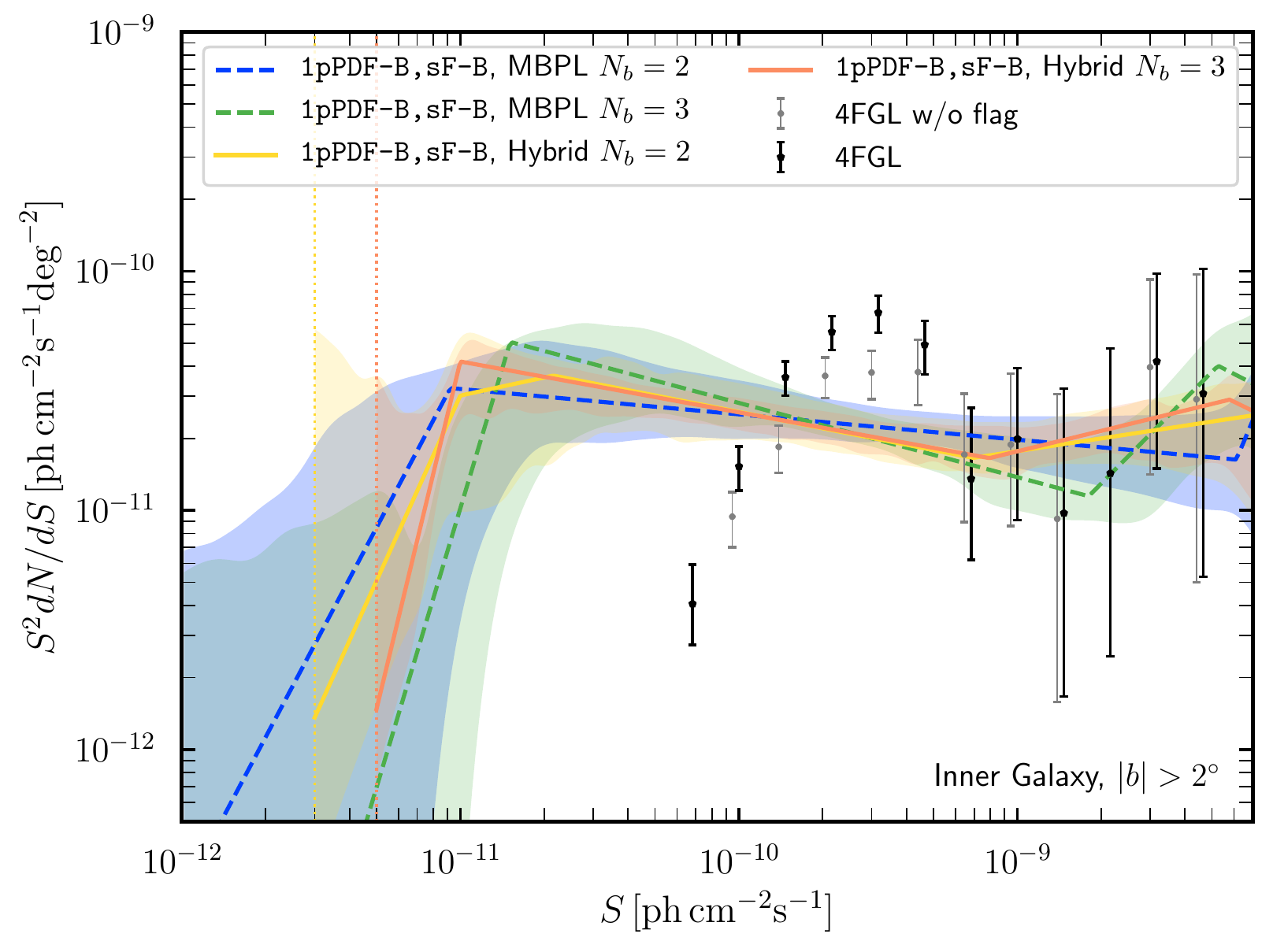}
  \includegraphics[width=0.48\textwidth]{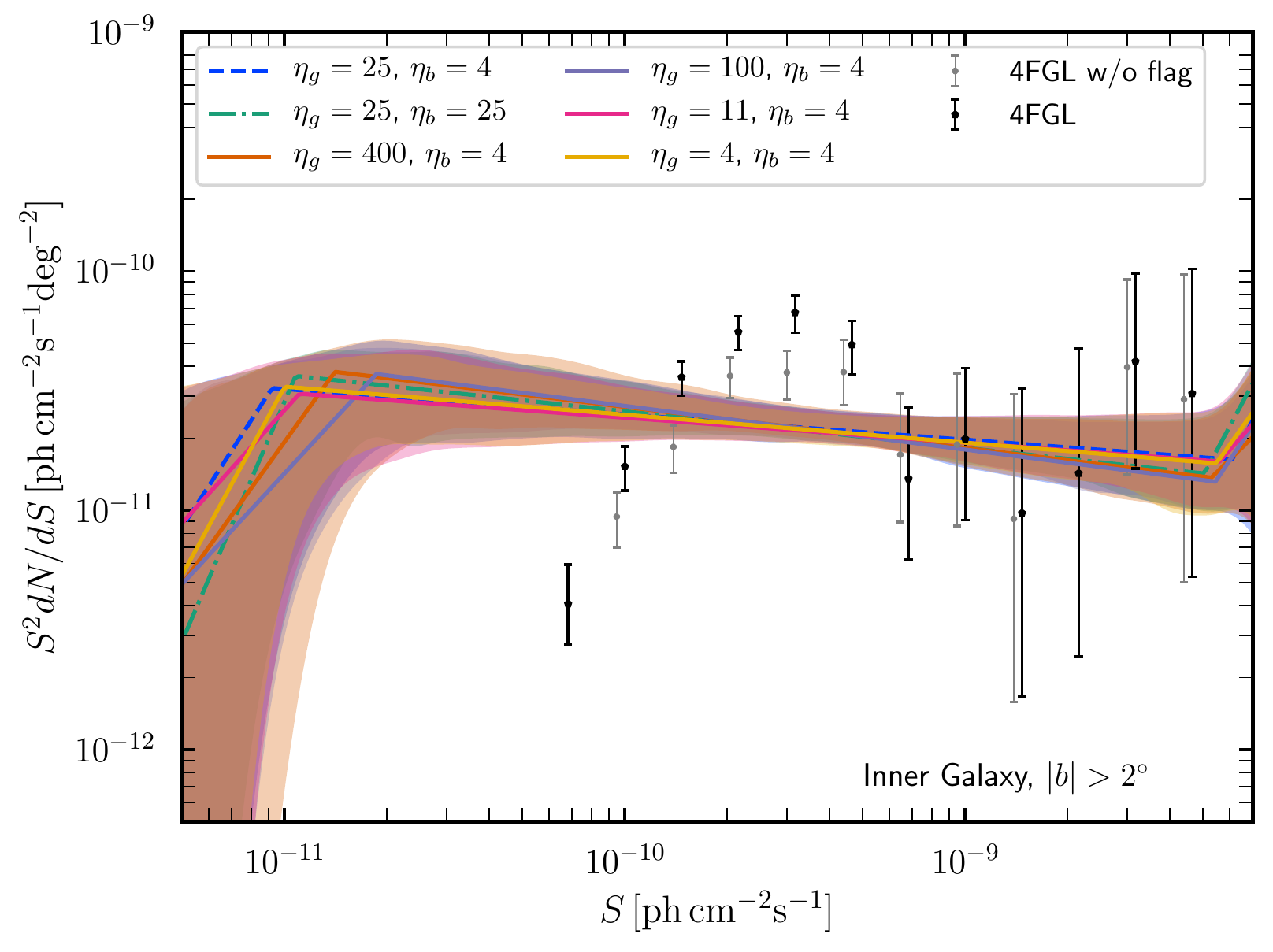}
  \caption{\textit{Systematic for \dnds~reconstruction in the IG.} Source count distribution of the IG obtained from the 1pPDF analysis cutting the inner 2$^\circ$. Left panel: Effect of the number of free breaks $N_{\rm b}$ (dashed lines) and of the Hybrid fit approach with different number of breaks and varying the node position. The dotted line illustrates the position of $S_{\rm nd1}$ for the corresponding Hybrid fit. 
Right panel: Effect of the smoothing scales $\eta_{\rm g,b}$ used to obtain the \SF~diffuse emission. 
  Points as in Fig.~\ref{fig:GDEmodel_syst}.}
  \label{fig:smoothing}
\end{figure*}

The \dnds~in the IG and OG is well described by a MBPL with two free breaks. We verified that an additional free break is not preferred by data, and that the MBPL obtained with three free breaks (see prior intervals in Table \ref{tab:priors}) is compatible, within the uncertainties, with the case of two free breaks. This is illustrated in the left panel of Fig.~\ref{fig:smoothing}. 

Fluxes from bright sources  with $S>10^{-8}$ ph cm$^{-2}$ s$^{-1}$ are not considered in the \Opp \,analysis. We verified that shifting the upper flux cut down to $S>10^{-9}$ ph cm$^{-2}$ s$^{-1}$ does not change our main results. The only difference we observe is a change in the position of the first flux break, driven by few, bright sources, always compatible with 4FGL points and uncertainties within $1-2\sigma$ C.L.
As for the low-flux regime, in our benchmark \Opp \,setup we do not include any lower flux cut. 
To test for possible effects connected to the faint end of the source-count distribution, we repeated the main analysis using the hybrid approach introduced in \cite{Zechlin:1}.
We set a fixed node at $S_{\rm nd1}$, with the index of the power-law component below the last node, $n_f=-10$, thus effectively suppressing possible contributions in the ultra-faint regime below the fixed node. 
Note that a fixed node $S_{\rm nb0}$ at the lower limit of the prior for the last free break is technically imposed, since  the first free node $S_{\rm nb1}$ is continuously
connected to the MBPL component with a power law 
at higher fluxes. 
We tested different values for the position of $S_{\rm nd1}$ in the faint source regime, together with a MBPL with two or three free breaks. 
Results are summarized in the left panel of Fig.~\ref{fig:smoothing}. 
To the extent we have tested, a node in the faint source regime at $3-5\cdot 10^{-12}$ \fluxunits does not affect the reconstructed \dnds~of the IG, which is well compatible, within $1\sigma$ uncertainty bands, with the benchmark results discussed in Fig.~1. In particular, the \dnds~is well compatible in the flux interval $10^{-11}-10^{-9}$ \fluxunits, where the radial and longitude profiles are computed.

We also tested the effect on the \Opp~results of changing the smoothing scale $\eta$ of the Galactic diffuse components in the \SF~fit.
The benchmark values used for the results illustrated so far are $\eta_{\rm g}=25$ and $\eta_{\rm b}=4$ for the gas and the \Fermi~bubbles components, respectively. 
Variations for $\eta_{\rm b}=25$ and different values for $\eta_{\rm g}=400,100,11,4$ are tested to assess possible systematic connected to this choice. 
In fact, a higher value of $\eta$ corresponds to a higher smoothing in the \SF~template used for \Opp~analysis. A different smoothing scale in the \SF~diffuse template could affect the reconstructed \dnds~in the \Opp, as this could leave residuals at small angular scales that could be wrongly attributed to PS by the \Opp. 
In Fig.~\ref{fig:smoothing} the corresponding results for the \dnds~are reported, compared to the benchmark results also shown in Fig.~1.
This test is performed using the \SFBBn~diffuse emission within  the \OppB~fit.
The best fit and $1\sigma$ band of the reconstructed source-count distribution are consistent for all the explored variations in $\eta_{\rm b,g}$. The same is true for the best-fit parameters of the IPS. 
As for the other \Opp~parameters, variations of $\eta_{\rm b}$ do not affect significantly the best fit of $A_B$ and the $\ln(\mathcal{Z})$.
We instead observe a slight decrease (increase) of $\ln(\mathcal{Z})$ for decreasing (increasing) $\eta_{\rm g}$. This is expected, as for smaller $\eta_{\rm g}$  in the \SF~fit  the Galactic diffuse emission gas template can  more easily account for small variations between adjacent pixels, decreasing the overall residuals. 

Finally, we note that, when analyzing the inner Galaxy divided in rings as shown in Fig.\ref{fig:slices}, the best-fit normalization  for $A_B$   decreases slightly going towards the innermost ring. However,  the uncertainties get larger, making the bulge template normalization $A_B$  in the rings still compatible with the numbers shown in Table I within $1\sigma$ uncertainties.

\section{Outer Galaxy and extragalactic \dnds~results}
\label{sm:ROI}
\begin{figure*}[t]
  \includegraphics[width=0.48\textwidth]{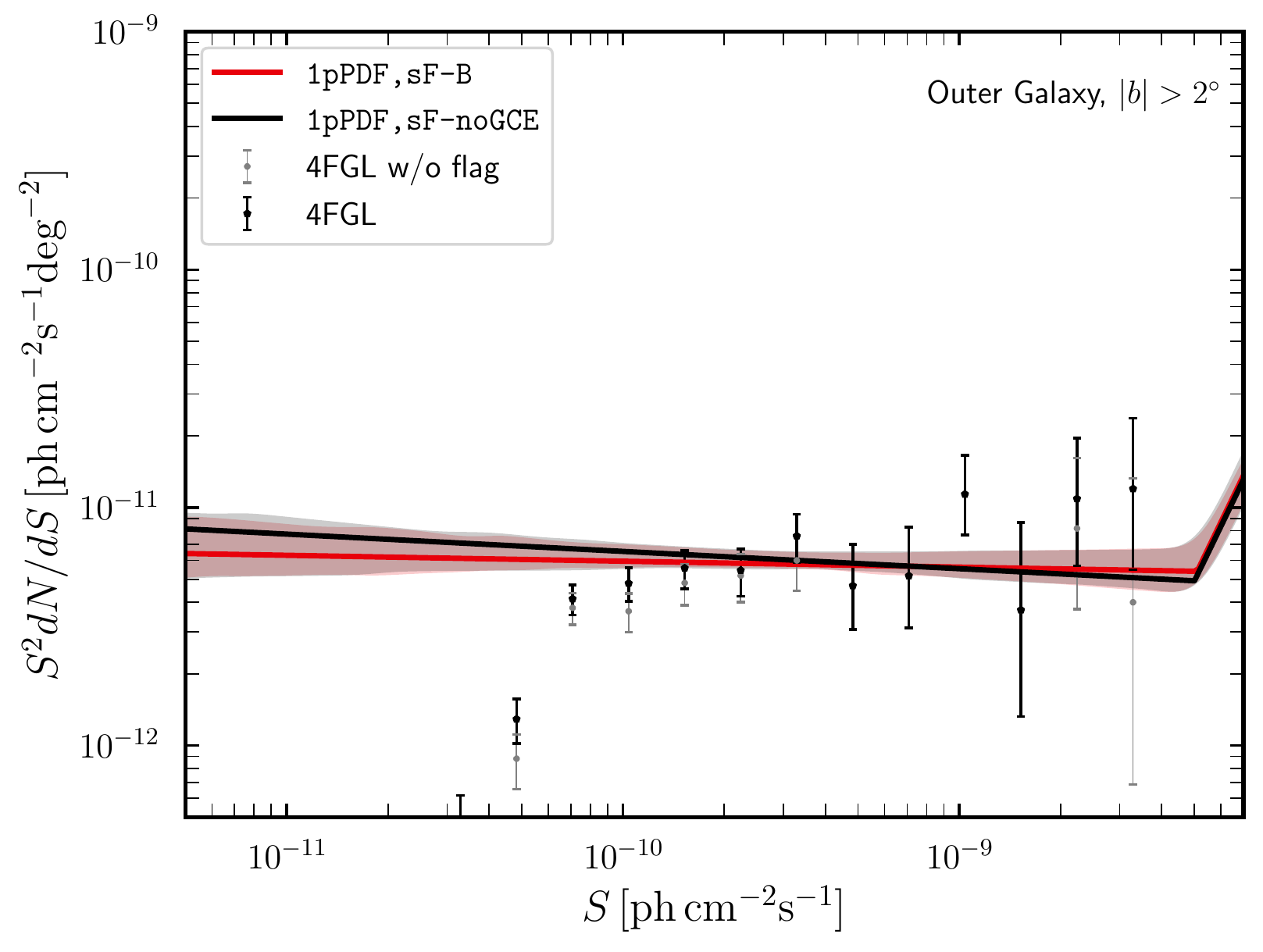}
    \includegraphics[width=0.48\textwidth]{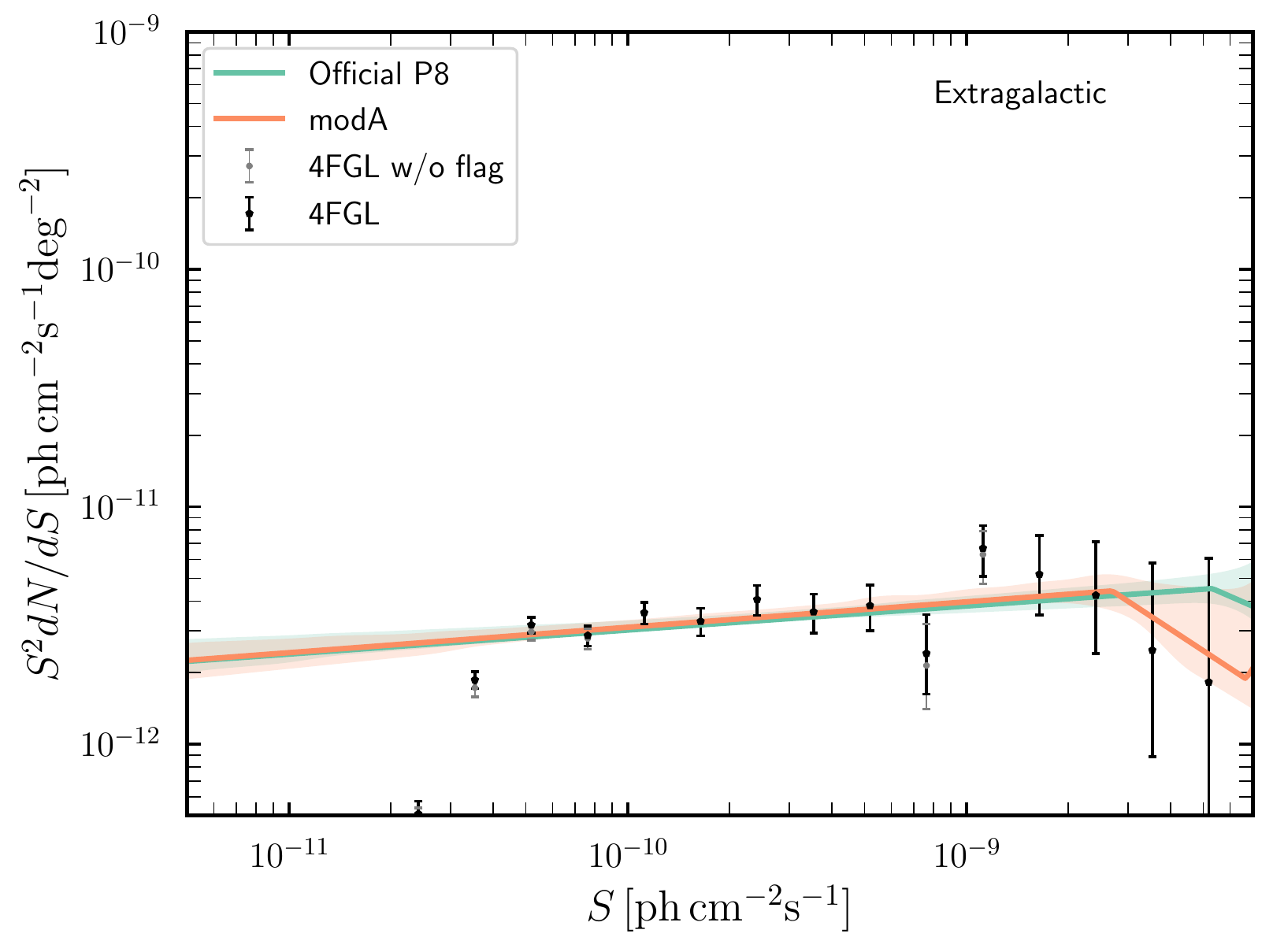}
\caption{\textit{Outer Galaxy and extragalactic \dnds.} Source count distribution of the outer Galaxy (left panel) and of the extragalactic ROI (right panel) obtained from the \Opp~analysis. 
Outer Galaxy: Results for the \dnds~are reported for two cases: (i) using the \SF~Galactic diffuse emission model without any component modeling the GCE (black line), and for the benchmark \SFBBn~model, where an additional component modeling the bulge is added. 
Extragalactic: Results are here reported using the Official P8 model and modA. 
Points as in Fig.~\ref{fig:GDEmodel_syst}.} 
  \label{fig:OGEG}
\end{figure*}
We finally report extended results on the OG and EG ROI, which are illustrated in Fig.~\ref{fig:OGEG}.
We note that in these ROI there are no 4FGL flagged sources, and the \dnds~is well described by a single power law from $S\sim3\cdot 10^{-9}$ \fluxunits down to $S \sim 10^{-11}$ ph~cm$^{-2}$s$^{-1}$. 
The IPS in the OG and EG are well consistent with 4FGL source counts for bright sources. 
In both cases, we obtain compatible results among different Galactic diffuse emission models, both for \SF~(left panel) or for the Official P8 and modA (right panel). 
Results for the [2, 5]~GeV  IPS in the EG  are also well compatible with previous studies \cite{Zechlin:2,Zechlin:3}.

\end{document}